\documentclass[a4paper,11pt]{article}
\usepackage[DIV12]{typearea}
\usepackage{longtable}
\usepackage{booktabs,colortbl}
\usepackage{multirow}
\usepackage{amsmath}
\usepackage{amssymb}
\usepackage{bbm}
\usepackage[utf8]{inputenc}
\usepackage{pdflscape}
\usepackage{xspace}
\usepackage[caption=false]{subfig}
\usepackage{nicefrac}
\usepackage{graphicx}
\usepackage{cite}  
\usepackage[small]{caption2}
\usepackage{mathtools}
\usepackage{nccfoots}

\newcommand{\rep}[1]{\ensuremath\boldsymbol{#1}}

\newcommand{\Z}[1]{\ensuremath{\mathbbm{Z}_{#1}}} 
\newcommand{\SO}[1]{\ensuremath{\mathrm{SO}(#1)}}
\newcommand{\SU}[1]{\ensuremath{\mathrm{SU}(#1)}}
\newcommand{\SL}[1]{\ensuremath{\mathrm{SL}(#1)}}
\newcommand{\U}[1]{\ensuremath{\mathrm{U}(#1)}}
\newcommand{\E}[1]{\ensuremath{\mathrm{E}_{#1}}}

\newcommand{\I}{\mathrm{i}}
\newcommand{\Id}{\mathbbm{1}}

\newcommand{\CP}{\ensuremath{\mathcal{CP}}\xspace}
\newcommand{\x}{\ensuremath{\times}}

\newcommand{\vequal}{\rotatebox[origin=c]{-90}{$=$}}

\newcommand{\W}[2]{\ensuremath{\hat{W}\!\left(\!\begin{smallmatrix}#1\\ #2\end{smallmatrix}\!\right)}}
\newcommand{\M}[2]{\ensuremath{M\!\left(\!\begin{smallmatrix}#1\\ #2\end{smallmatrix}\!\right)}}
\newcommand{\Sp}[1]{\ensuremath{\mathrm{Sp}(#1,\mathbbm{Z})}}

\usepackage{xcolor}
\definecolor{darkgreen}{HTML}{109930}
\definecolor{pink}{rgb}{0.858, 0.188, 0.478}

\advance \headheight by 3.0truept       
\usepackage[pdftex,hidelinks]{hyperref}
\hypersetup{
    pdftitle = {Orbifolds from Sp(4,Z) and their modular symmetries},
    pdfauthor = {Nilles, Ramos-Sanchez, Trautner, Vaudrevange}
}

\begin{document}

\begin{titlepage}

\begin{flushright}
\normalsize{TUM-HEP 1339/21}
\end{flushright}

\vspace*{1.0cm}

\begin{center}
{\Large\textbf{\boldmath Orbifolds from $\mathrm{Sp}(4,\Z{})$ and their modular symmetries}\unboldmath}

\vspace{1cm}

\textbf{Hans~Peter~Nilles}$^{a}$,
\textbf{Sa\'ul Ramos--S\'anchez}$^{b}$, \textbf{Andreas Trautner}$^{c}$, \\ and \textbf{Patrick~K.S.~Vaudrevange}$^{d}$
\Footnote{*}{%
\href{mailto:nilles@th.physik.uni-bonn.de;ramos@fisica.unam.mx;trautner@mpi-hd.mpg.de;patrick.vaudrevange@tum.de}{\tt Electronic addresses} 
}
\\[5mm]
\textit{$^a$\small Bethe Center for Theoretical Physics and Physikalisches Institut der Universit\"at Bonn,\\ Nussallee 12, 53115 Bonn, Germany}
\\[2mm]
\textit{$^b$\small Instituto de F\'isica, Universidad Nacional Aut\'onoma de M\'exico,\\ POB 20-364, Cd.Mx. 01000, M\'exico}
\\[2mm]
\textit{$^c$\small Max-Planck-Institut f\"ur Kernphysik, \\ Saupfercheckweg 1, 69117 Heidelberg, Germany}
\\[2mm]
\textit{$^d$\small Physik Department T75, Technische Universit\"at M\"unchen,\\ James-Franck-Stra\ss e 1, 85748 Garching, Germany}
\end{center}

\vspace{1cm}

\vspace*{1.0cm}

\begin{abstract}
The incorporation of Wilson lines leads to an extension of the modular symmetries of string 
compactification beyond \SL{2,\Z{}}. In the simplest case with one Wilson line $Z$, K\"ahler 
modulus $T$ and complex structure modulus $U$, we are led to the Siegel modular group \Sp{4}. 
It includes $\SL{2,\Z{}}_T\times\SL{2,\Z{}}_U$ as well as \Z2 mirror symmetry, which 
interchanges $T$ and $U$. Possible applications to flavor physics of the Standard Model 
require the study of orbifolds of \Sp{4} to obtain chiral fermions. We identify the 13 
possible orbifolds and determine their modular flavor symmetries as subgroups of \Sp{4}. 
Some cases correspond to symmetric orbifolds that extend previously discussed cases of 
\SL{2,\Z{}}. Others are based on asymmetric orbifold twists (including mirror symmetry) 
that do no longer allow for a simple intuitive geometrical interpretation and require 
further study. Sometimes they can be mapped back to symmetric orbifolds with quantized 
Wilson lines. The symmetries of \Sp{4} reveal exciting new aspects of modular symmetries 
with promising applications to flavor model building.
\end{abstract}

\end{titlepage}

\newpage

\section{Introduction}

Modular symmetries appear frequently in string theory. They might have applications in particle 
physics as discrete non-Abelian flavor symmetries. In the simplest case, these discrete modular 
symmetries descend from the modular group $\SL{2,\Z{}}$ of a two-dimensional torus, on which two 
extra spatial dimensions have been compactified. The implementation within string theory requires 
some aspects of model building towards the $\SU{3}\times\SU{2}\times\U{1}$ Standard Model of 
particle physics. One of the key aspects is the desired presence of chiral fermions. This requires 
a twist of the torus. Then, chiral matter fields can be realized in the twisted sectors of the 
orbifold, located at the ``fixed points'' of the orbifold twist. Simplest examples correspond to 
the $\Z{K}$ orbifolds $\mathbbm{T}^2/\Z{K}$ ($K=2,3,4,6$), where a full analysis has been performed 
recently~\cite{Nilles:2020tdp,Nilles:2020gvu,Baur:2020jwc,Baur:2021mtl}. They would correspond to 
six-dimensional string compactifications with an elliptic fibration.

In fact, in string theory a two-torus with background $B$-field is described by two moduli, the 
K\"ahler modulus $T$ and the complex structure modulus $U$ with modular symmetries 
$\SL{2,\Z{}}_T\times\SL{2,\Z{}}_U$. This is accompanied by mirror symmetry which interchanges $T$ 
and $U$. In the $\Z{K}$ orbifolds with ($K>2$), the complex structure modulus is frozen to allow 
for the orbifold twist and in these cases we have one unconstrained modulus $T$. This is, however, 
not the case for the \Z2 orbifold, where we remain with two unconstrained moduli and manifest 
mirror symmetry.

In general, string theories have a much richer moduli structure as they require the 
compactification of six spatial dimensions. But even if we concentrate on a two-dimensional 
subsector, we have additional moduli in the form of gauge background fields (Wilson lines). The 
moduli structure of the simplest example of such a system is given by the Siegel modular group 
$\mathrm{Sp}(4,\Z{})$ with moduli $T$, $U$ and one additional Wilson line modulus $Z$.
This can be made manifest in the Narain lattice formulation~\cite{Baur:2020yjl}. In addition,
this work is motivated by the recent bottom-up consideration of $\mathrm{Sp}(4,\Z{})$ flavor 
symmetries, see refs.~\cite{Ding:2020zxw,Ding:2021iqp}.

$\mathrm{Sp}(4,\Z{})$ contains as subgroups $\SL{2,\Z{}}_T\x\SL{2,\Z{}}_U$ as well as mirror 
symmetry. From the string theory perspective, an application of $\mathrm{Sp}(4,\Z{})$ as modular 
flavor symmetry would again require some orbifolding to obtain chiral fermions. The first step in 
this direction is a classification of orbifolds from $\mathrm{Sp}(4,\Z{})$, which is the main
purpose of the present paper. This is a generalization of the $\Z{K}$ orbifolds mentioned earlier. 
To perform the classification of $\mathrm{Sp}(4,\Z{})$ orbifolds, we realize that each inequivalent 
fixed point in the string moduli space $(T,U,Z)$ that is left invariant by a subgroup of 
$\mathrm{Sp}(4,\Z{})$ corresponds to an inequivalent orbifold (or, in general, to a set of 
orbifolds). Hence, the fixed points of $\mathrm{Sp}(4,\Z{})$ correspond to string orbifolds. Then, 
chiral fermions could appear at the fixed points of these $\mathrm{Sp}(4,\Z{})$ orbifold actions on 
the extra-dimensional space.\footnote{Hence, there are two different kinds of fixed points: First, 
the fixed points of $\mathrm{Sp}(4,\Z{})$ acting on the moduli space $(T,U,Z)$ and, second, the 
fixed points of the orbifold action in extra-dimensional space, where chiral fermions can be 
localized.} A classification of the fixed points of $\mathrm{Sp}(4,\Z{})$ has been given by 
Gottschling~\cite{Gottschling:1961fp,Gottschling:1961un,Gottschling:1967dg} long ago. There are 
altogether 13 different cases: two with complex dimension 2, five with complex dimension 1 and six 
of dimension 0.

The next step is the construction of those orbifolds that stabilize these fixed loci in moduli 
space. Our results are summarized in 
table~\ref{tab:OrbifoldSummary}. We identify the conventional (geometrical) twists on the moduli 
and, as a new mechanism, twists via mirror symmetry. As a result of this, asymmetric orbifolds 
appear frequently (although some of them are dual to symmetric orbifolds with specifically 
transformed moduli). A direct intuitive geometrical interpretation is often not available as the 
presence of Wilson lines and asymmetric twists introduce some ``non-geometrical'' aspects.

If we set the Wilson lines to zero, we obtain the $\mathbbm{T}^2/\Z{K}$ examples discussed earlier: 
Section~\ref{sec:Z2Symmetric} in table~\ref{tab:OrbifoldSummary} represents the symmetric \Z2 
orbifold, section~\ref{sec:Z4Symmetric} the \Z4 orbifold, and section~\ref{sec:Z6Symmetric} the 
symmetric \Z3 orbifold (embedded in the \Z6 case). The orbifold of section~\ref{sec:Z2Asymmetric} 
corresponds to an asymmetric \Z2 orbifold. In this case, however, we can define a duality 
transformation that maps it to a symmetric \Z2 orbifold with a quantized Wilson line (discussed 
in section~\ref{sec:Z2SymmetricWL}).

\enlargethispage{0.4cm}
The paper is organized as follows. In section~\ref{sec:OrbifoldsAndSp4Z}, we introduce the modular 
transformations of $\mathrm{Sp}(4,\Z{})$ on the string moduli (see eq.~\eqref{eq:Sp4TmodTrafos}) 
within the framework of Narain orbifold compactifications. Section~\ref{sec:StabilizedModuliByOrb} 
discusses case by case the stabilization of moduli by \Sp{4} orbifolds that lead to 
the results summarized in table~\ref{tab:OrbifoldSummary}. In addition, we explicitly construct for 
each orbifold the unbroken modular group $\mathcal{G}_\mathrm{modular}$ as a subgroup of 
$\mathrm{Sp}(4,\Z{})$ plus a \CP-like transformation. While these are important steps towards 
applications to the flavor problem of the Standard Model of particle physics, there are still many 
open questions. These will be mentioned in section~\ref{sec:Conclusions}, devoted to conclusions 
and outlook.

\begin{table}[t!]
\centering
\begin{tabular}{c|c|c|c|c}
\toprule
complex dimension & Narain point group    & type        & moduli    & reference to\\
of moduli space   & of orbifold           & of orbifold & $(T,U,Z)$ & section\\
\midrule
\arrayrulecolor{lightgray}
2            & $\Z{2}$                    & symmetric   & $(T,U,0)$                                           & \ref{sec:Z2Symmetric}\\[-2pt]
2            & $\Z{2}$                    & asymmetric  & $(T,T,Z)$                                           & \ref{sec:Z2Asymmetric} \\[-2pt]
2            & $\Z{2}$                    & $\begin{array}{c}\mathrm{symmetric}\\[-0.2cm]\mathrm{(dual\ to\ \ref{sec:Z2Asymmetric})}\end{array}$  & $(T,U,\nicefrac{1}{2})$ & \ref{sec:Z2SymmetricWL}\\[-2pt]
\cmidrule{1-5} 
1            & $\Z{4}$                    & symmetric   & $(T,\I,0)$                                          & \ref{sec:Z4Symmetric}\\[-2pt]
1            & $\Z{6}$                    & symmetric   & $(T,\omega,0)$                                      & \ref{sec:Z6Symmetric}\\[-2pt]
1            & $\Z{2}\times\Z{2}$         & asymmetric  & $(T,T,0)$                                           & \ref{sec:Z2xZ2Asymmetric}\\[-2pt]
1            & $\Z{2}\times\Z{2}$         & asymmetric  & $(T,T,\nicefrac{1}{2})$                             & \ref{sec:Z2xZ2AsymmetricWL}\\[-2pt]
1            & $S_3$                      & asymmetric  & $(T,T,\nicefrac{T}{2})$                             & \ref{sec:S3IAsymmetric}\\[-2pt]
\cmidrule{1-5} 
0            & $\Z{5}$                    & asymmetric  & $(-\zeta^{-1}, \zeta, \zeta+\zeta^{-2})$            & \ref{sec:Z5Asymmetric}\\[-2pt]
0            & $S_4$                      & asymmetric  & $(\tilde\eta,\tilde\eta,\frac{1}{2}(\tilde\eta-1))$ & \ref{sec:S4Asymmetric}\\[-2pt]
0            & $(\Z4\times\Z2)\rtimes\Z2$ & asymmetric  & $(\I,\I,0)$                                         & \ref{sec:Z4Z2Z2Asymmetric}\\[-2pt]
0            & $S_3\times\Z{6}$           & asymmetric  & $(\omega,\omega,0)$                                 & \ref{sec:S3xZ6Asymmetric}\\[-2pt]
0            &$S_3\times\Z{2}\cong D_{12}$& asymmetric  & $\frac{\I}{\sqrt{3}}(2,2,1)$                        & \ref{sec:S3IIAsymmetric}\\[-2pt]
0            & $\Z{12}$                   & asymmetric  & $(\I,\omega,0)$                                     & \ref{sec:Z12Asymmetric}\\[-2pt]
\arrayrulecolor{black}
\bottomrule
\end{tabular}
\vspace{-0.4cm}
\caption{Summary of symmetric and asymmetric orbifold compactifications with Narain point 
groups classified by the inequivalent fixed points $(T,U,Z)$ of \Sp{4}. We use the 
definitions $\omega :=\exp(\nicefrac{2\pi\I}{3})$, $\zeta:=\exp\left(\nicefrac{2\pi\I}{5}\right)$ 
and $\tilde\eta:=\frac{1}{3}(1+2\,\sqrt{2}\,\I)$.}
\label{tab:OrbifoldSummary}
\end{table}

\newpage

\section[(A)symmetric orbifolds and Sp(4,Z)]{\boldmath (A)symmetric orbifolds and $\mathrm{Sp}(4,\Z{})$\unboldmath}
\label{sec:OrbifoldsAndSp4Z}

\subsection{\boldmath Narain torus compactification and string moduli\unboldmath}

To construct an orbifold in the Narain formulation of the heterotic string, we first discuss a 
general $D$-dimensional torus compactification with $B$-field and Wilson line 
backgrounds~\cite{Narain:1985jj,Narain:1986am}. To do so, we have to impose torus boundary 
conditions on the $D$ right- and $D+16$ left-moving (bosonic) string modes 
$(y_\mathrm{R}, y_\mathrm{L})$, respectively,
\begin{equation}\label{eq:NarainBC}
\begin{pmatrix}y_\mathrm{R}\\y_\mathrm{L}\end{pmatrix} ~\sim~ \begin{pmatrix}y_\mathrm{R}\\y_\mathrm{L}\end{pmatrix} + E\,\hat{N}\;,\quad\mathrm{where}\quad \hat{N} ~=~ \begin{pmatrix}n\\m\\p\end{pmatrix} ~\in~\Z{}^{2D+16}\;.
\end{equation}
The $(2D+16)$-dimensional vector of integers $\hat{N}$ contains the winding numbers $n\in\Z{}^D$, 
the Kaluza--Klein numbers $m\in\Z{}^D$, and the gauge quantum numbers $p\in\Z{}^{16}$ corresponding 
to the $\E{8}\times\E{8}$ (or $\SO{32}$) gauge group of the supersymmetric heterotic string. 
Furthermore, $E$ denotes the Narain vielbein. For the worldsheet one-loop string vacuum amplitude 
to be modular invariant, $E$ has to satisfy
\begin{equation}
E^\mathrm{T}\, \eta\, E ~=~ \hat\eta ~:=~ \begin{pmatrix}0&\Id_D&0\\\Id_D&0&0\\0&0&g\end{pmatrix}\;, \quad\mathrm{where}\quad \eta ~:=~ \begin{pmatrix}-\Id_D&0&0\\0&\Id_D&0\\0&0&\Id_{16}\end{pmatrix}\;.
\end{equation}
Here, $\eta$ is the Narain metric of signature $(D,D+16)$, $g:=\alpha_\mathrm{g}^\mathrm{T}\alpha_\mathrm{g}$ 
and the columns of the $(16 \times 16)$-dimensional matrix $\alpha_\mathrm{g}$ contain the simple 
roots of $\E{8}\times\E{8}$ (or $\mathrm{Spin}(32)/\Z{2}$). Consequently, $E$ spans a so-called 
Narain lattice: an even, integer, self-dual lattice of signature $(D,D+16)$. The ``rotational'' 
symmetries of the Narain lattice give rise to the so-called modular group of the Narain lattice, 
\begin{equation}
\mathrm{O}_{\hat{\eta}}(D,D+16,\Z{}) ~:=~ \big\langle ~\hat\Sigma~\big|~ \hat\Sigma ~\in~\mathrm{GL}(2D+16,\Z{}) 
\quad\mathrm{with}\quad \hat\Sigma^\mathrm{T}\hat\eta\,\hat\Sigma = \hat\eta~\big\rangle\;.
\end{equation}
In order to understand the action of $\mathrm{O}_{\hat{\eta}}(D,D+16,\Z{})$, it is convenient to 
define the generalized metric of the Narain lattice,\footnote{We use the conventions of 
refs.~\cite{GrootNibbelink:2017usl,GrootNibbelink:2020dib} with $B$ replaced by $-B$.}
\begin{equation}\label{eq:GeneralizedMetric}
\mathcal{H} := E^\mathrm{T}E = \begin{pmatrix}
\frac{1}{\alpha'}\left(G+\alpha' A^\mathrm{T} A+C^\mathrm{T} G^{-1} C\right) & -C^\mathrm{T} G^{-1}                           & (\Id_D + C^\mathrm{T} G^{-1}) A^\mathrm{T} \alpha_\mathrm{g} \\
-G^{-1} C                                                                    & \alpha' G^{-1}                                 & - \alpha' G^{-1} A^\mathrm{T} \alpha_\mathrm{g}\\
\alpha_\mathrm{g}^\mathrm{T} A (\Id_D + G^{-1} C)                            & -\alpha' \alpha_\mathrm{g}^\mathrm{T} A G^{-1} & \alpha_\mathrm{g}^\mathrm{T} \big(\Id_{16} + \alpha' A G^{-1} A^\mathrm{T} \big) \alpha_\mathrm{g}
\end{pmatrix}
\end{equation}
and $C := B + \frac{\alpha'}{2} A^\mathrm{T} A$. Here, $\alpha'$ is the Regge slope that renders 
$E$ dimensionless, $G=e^{\mathrm{T}} e$ is the metric of the $D$-dimensional torus $\mathbbm{T}^D$ 
defined by the torus basis vectors contained as the columns of the $D \times D$ vielbein matrix 
$e$. In addition, $B$ is the anti-symmetric background $B$-field, while the $16\times D$ matrix $A$ 
gives rise to the Wilson lines along the $D$ torus directions.

In the following, we will mainly set $D=2$ and choose the two column vectors $A_i$ of the Wilson 
line matrix $A$ as
\begin{equation}
A_{i} ~=~ (a_i, -a_i, 0, \ldots, 0)^\mathrm{T} \quad\mathrm{with}\quad a_i ~\in~ \mathbbm{R} \quad\mathrm{for}\quad i~\in~\{1,2\}\;.
\end{equation}
Then, we define the (dimensionless) string moduli as~\cite{Mayr:1995rx}
\begin{subequations}
\begin{eqnarray}
T & := & \frac{1}{\alpha'}\left(B_{12}+\I\,\sqrt{\mathrm{det}\,G}\right)+a_1\left(-a_2+U\,a_1\right)\:,\\
U & := & \frac{1}{G_{11}}\left(G_{12}+\I\,\sqrt{\mathrm{det}\,G}\right)\;,\\
Z & := & -a_2+U\, a_1\;.
\end{eqnarray}
\end{subequations}
The K\"ahler modulus $T$ determines the $B$-field background, the overall size of the 
extra-dimensional two-torus, and is altered by the Wilson line parameters $a_i$. The complex 
structure modulus $U$ parameterizes the shape of the two-torus $\mathbbm{T}^2$, while the Wilson 
line modulus $Z$ depends on the parameters $a_i$. The transformation of these moduli under a 
modular transformation $\hat\Sigma \in \mathrm{O}_{\hat{\eta}}(2,2+16,\Z{})$ can be computed by 
considering the generalized metric
\begin{equation}\label{eq:TrafoOfModuli}
\mathcal{H}(T,U,Z) ~\xmapsto{~\hat\Sigma~}~ \mathcal{H}(T',U',Z') ~:=~ \hat\Sigma^{-\mathrm{T}}\mathcal{H}(T,U,Z)\, \hat\Sigma^{-1}\;,
\end{equation}
see for example ref.~\cite{Baur:2019iai}. For the transformations 
$\hat\Sigma\in\{\hat{K}_\mathrm{S},\,\hat{K}_\mathrm{T},\,\hat{C}_\mathrm{S},\,\hat{C}_\mathrm{T},\,\hat{M},\,\W{\ell}{m},\,\hat\Sigma_*\}$ 
given in ref.~\cite{Baur:2020yjl} this yields
\begin{subequations}\label{eq:Sp4TmodTrafos}
\begin{align}
T &~\xmapsto{~\hat{K}_\mathrm{S}~}~ -\frac{1}{T}\;,    & U &~\xmapsto{~\hat{K}_\mathrm{S}~}~ U-\frac{Z^2}{T}\;, & Z &~\xmapsto{~\hat{K}_\mathrm{S}~}~ -\frac{Z}{T}\;,\label{eq:ModuliTrafoKS}\\
T &~\xmapsto{~\hat{K}_\mathrm{T}~}~ T+1\;,             & U &~\xmapsto{~\hat{K}_\mathrm{T}~}~ U\;,               & Z &~\xmapsto{~\hat{K}_\mathrm{T}~}~ Z\;,\label{eq:ModuliTrafoKT}\\
T &~\xmapsto{~\hat{C}_\mathrm{S}~}~ T-\frac{Z^2}{U}\;, & U &~\xmapsto{~\hat{C}_\mathrm{S}~}~-\frac{1}{U}\;,     & Z &~\xmapsto{~\hat{C}_\mathrm{S}~}~-\frac{Z}{U}\;,\label{eq:ModuliTrafoCS}\\
T &~\xmapsto{~\hat{C}_\mathrm{T}~}~ T\;,               & U &~\xmapsto{~\hat{C}_\mathrm{T}~}~ U+1\;,             & Z &~\xmapsto{~\hat{C}_\mathrm{T}~}~ Z\;,\label{eq:ModuliTrafoCT}\\
T &~\xmapsto{~\hat{M}~}~ U\;,                          & U &~\xmapsto{~\hat{M}~}~ T\;,                          & Z &~\xmapsto{~\hat{M}~}~ Z\;,\label{eq:ModuliTrafoMirror}\\
T &~\xmapsto{\W{\ell}{m}}~ T+m(m\,U+2\,Z-\ell)\;,      & U &~\xmapsto{\W{\ell}{m}}~ U\;,                        & Z &~\xmapsto{\W{\ell}{m}}~ Z+m\,U-\ell\;,\\
T &~\xmapsto{~\hat\Sigma_*~}~ -\bar{T}\;,              & U &~\xmapsto{~\hat\Sigma_*~}~ -\bar{U}\;,              & Z &~\xmapsto{~\hat\Sigma_*~}~ -\bar{Z}\;.
\end{align}
\end{subequations}

\subsection{\boldmath Narain orbifold compactification and string moduli\unboldmath}

We can extend the Narain torus boundary conditions~\eqref{eq:NarainBC} by an orbifold 
action~\cite{Dixon:1985jw,Dixon:1986jc,Ibanez:1986tp,Narain:1986qm,Narain:1990mw,GrootNibbelink:2017usl,GrootNibbelink:2020dib} 
\begin{equation}\label{eq:NarainOrbifoldBC}
\begin{pmatrix}y_\mathrm{R}\\y_\mathrm{L}\end{pmatrix} ~\sim~ \Theta\, \begin{pmatrix}y_\mathrm{R}\\y_\mathrm{L}\end{pmatrix} + E\,\hat{N}\;,\quad\mathrm{where}\quad \Theta~:=~ \begin{pmatrix}\theta_\mathrm{R}&0\\0&\theta_\mathrm{L}\end{pmatrix} ~\in~ \mathrm{O}(D) \oplus \mathrm{O}(D+16)\;,
\end{equation}
and $\hat{N}\in\Z{}^{2D+16}$. The rotation matrix $\Theta$ denotes the so-called Narain twist. The 
set of all Narain twists generates the so-called Narain point group $P_\mathrm{Narain}$. If there 
exists a basis such that $\theta_\mathrm{L} = \theta_\mathrm{R} \oplus \Id_{16}$ for all twists, 
the orbifold is called symmetric, and otherwise it is called asymmetric. Moreover, the orbifold 
action~\eqref{eq:NarainOrbifoldBC} suggests to define transformations $(\Theta, E\,\hat{N})$, which 
generate the so-called Narain space group $S_\mathrm{Narain}$. Then, an orbifold compactification 
(of worldsheet bosons) is fully specified by the choice of $S_\mathrm{Narain}$. Due to its 
right-left structure $\theta_\mathrm{R}$ and $\theta_\mathrm{L}$ in eq.~\eqref{eq:NarainOrbifoldBC}, 
a Narain twist $\Theta$ has to satisfy the conditions
\begin{equation}\label{eq:ConditionsOnTwist}
\Theta^\mathrm{T}\,\Theta ~=~ \Id_{2D+16} \qquad\mathrm{and}\qquad \Theta^\mathrm{T}\,\eta\,\Theta ~=~ \eta\;.
\end{equation}
Furthermore, the Narain twist has to map the Narain lattice to itself. Hence, it is convenient to 
define the Narain twist in the Narain lattice basis
\begin{equation}
\hat\Theta ~:=~ E^{-1}\, \Theta\, E ~\in~ \mathrm{GL}(2D+16,\Z{})\;,
\end{equation}
such that $\hat\Theta$ is an integer matrix from $\mathrm{GL}(2D+16,\Z{})$. In the Narain lattice 
basis, we denote the Narain point group by $\hat{P}_\mathrm{Narain}$ and the Narain space group by 
$\hat{S}_\mathrm{Narain}$, where its elements are of the form $(\hat\Theta, \hat{N})$. Using the 
Narain lattice basis, the conditions~\eqref{eq:ConditionsOnTwist} read
\begin{equation}\label{eq:GeneralizedMetricInvariant}
\hat\Theta^\mathrm{T}\,\mathcal{H}\,\hat\Theta ~=~ \mathcal{H} \qquad\mathrm{and}\qquad \hat\Theta^\mathrm{T}\,\hat\eta\,\hat\Theta ~=~ \hat\eta\;.
\end{equation}
Consequently, $\hat\Theta\in\mathrm{O}_{\hat{\eta}}(D,D+16,\Z{})$ has to be an element of the 
modular group of the Narain lattice that leaves the generalized metric $\mathcal{H}$ invariant, see 
eq.~\eqref{eq:TrafoOfModuli}. In general, condition~\ref{eq:GeneralizedMetricInvariant} fixes some 
of the moduli. In other words, the Narain twist $\hat\Theta$ is a symmetry of the Narain lattice 
only for some special values of the moduli. In this case, we say that some moduli are stabilized 
geometrically by the orbifold action. We denote the modular group after orbifolding by 
$\mathcal{G}_\mathrm{modular}$. It is given by those elements $(\hat\Sigma, \hat{T})\not\in\hat{S}_\mathrm{Narain}$ 
with $\hat\Sigma\in\mathrm{O}_{\hat{\eta}}(D,D+16,\Z{})$ and $\hat{T}\in\mathbbm{Q}^{2D+16}$ that 
are outer automorphisms of the Narain space group $\hat{S}_\mathrm{Narain}$, cf.\ 
refs.~\cite{Baur:2019kwi,Baur:2019iai}. In the case where the Narain space group is generated only 
by elements of the form $(\hat\Theta, 0)$ and $(\Id_{2D+16}, \hat{N})$ with $\hat{N}\in\Z{}^{2D+16}$, 
the modular group after orbifolding is given by\footnote{If the Narain space group contains 
generators that are nontrivial roto-translations $(\hat\Theta, \hat{N})$ with 
$\hat{N}\not\in\Z{}^{2D+16}$, also the outer automorphisms of $\hat{S}_\mathrm{Narain}$ can be 
roto-translations, see e.g.\ appendix B in ref.~\cite{Baur:2020jwc}.}
\begin{equation}\label{eq:GModularOfOrbifold}
\mathcal{G}_\mathrm{modular} := \big\langle ~\hat\Sigma~\big|~ \hat\Sigma ~\in~\mathrm{O}_{\hat{\eta}}(D,D+16,\Z{})
\ \mathrm{with}\ \hat\Sigma^{-1} \hat\Theta\,\hat\Sigma \in \hat{P}_\mathrm{Narain} \ \mathrm{for\ all}\ \hat\Theta\in\hat{P}_\mathrm{Narain}~\big\rangle\;.
\end{equation}
Then, one can compute the transformation of the moduli after orbifolding using the generalized 
metric eq.~\eqref{eq:TrafoOfModuli}. 

In addition to the modular group $\mathcal{G}_\mathrm{modular}$ (which is in general an infinite, 
discrete group), there exists the closely related finite modular group $\mathcal{G}_\mathrm{fmg}$, 
which can play an important role in flavor physics~\cite{Feruglio:2017spp,Criado:2018thu,Feruglio:2019ktm}. 
This group appears in string theory as follows: On orbifolds, there are so-called twisted strings 
that are localized in extra dimensions at the fixed points of the orbifold action. They transform 
in general under a modular transformation $\hat\Sigma \in \mathcal{G}_\mathrm{modular}$ 
nontrivially with a unitary matrix representation $\rho_{\rep{r}}(\hat\Sigma)$ of a finite modular 
group $\mathcal{G}_\mathrm{fmg}$, for example $\mathcal{G}_\mathrm{fmg} \cong T'$ for the 
$\mathbbm{T}^2/\Z{3}$ orbifold~\cite{Lauer:1989ax,Lerche:1989cs,Lauer:1990tm,Baur:2019kwi,Baur:2019iai} 
and $\mathcal{G}_\mathrm{fmg} \cong (S_3^T \times S_3^U) \rtimes \Z{4}^{\hat{M}}$ for the 
$\mathbbm{T}^2/\Z{2}$ orbifold (without \CP)~\cite{Baur:2020jwc,Baur:2021mtl}. In addition, 
couplings $Y$ in the superpotential become modular forms of the moduli. Hence, couplings $Y$ also 
transform under a modular transformation $\hat\Sigma\in\mathcal{G}_\mathrm{modular}$ in a unitary 
matrix representation $\rho_Y(\hat\Sigma)$ of the finite modular group $\mathcal{G}_\mathrm{fmg}$. 
However, in some cases $\rho_Y$ is not a faithful representation of $\mathcal{G}_\mathrm{fmg}$. 
Then, one can compute the finite modular group $\mathcal{G}_\mathrm{fmg\ of\ Y}$ that is generated 
by the matrix representations $\rho_Y(\hat\Sigma)$ of the modular forms such that 
$\mathcal{G}_\mathrm{fmg\ of\ Y} \subset \mathcal{G}_\mathrm{fmg}$. For example, for the 
$\mathbbm{T}^2/\Z{2}$ orbifold we have $\mathcal{G}_\mathrm{fmg\ of\ Y} \cong (S_3^T \times S_3^U) \rtimes \Z{2}^{\hat{M}}$, 
see refs.~\cite{Baur:2020jwc,Baur:2021mtl}.

\noindent
In this paper we will frequently utilize the correspondences
\begin{equation}\label{eq:SP4ZBasisElements}
\begin{array}{cccccccc}
M_{(\mathrm{S},\Id_2)}\;,& M_{(\mathrm{T},\Id_2)}\;,& M_{(\Id_2,\mathrm{S})}\;,& M_{(\Id_2,\mathrm{T})}\;,& M_{\times}\;, &\M{\ell}{m}   & \mathrm{from} &\mathrm{Sp}(4,\Z{})\\[2pt]
\updownarrow             & \updownarrow             & \updownarrow             & \updownarrow             & \updownarrow  & \updownarrow & \\[2pt]
\hat{K}_\mathrm{S}\;,    & \hat{K}_\mathrm{T}\;,    & \hat{C}_\mathrm{S}\;,    & \hat{C}_\mathrm{T}\;,    & \hat{M}\;,    &\W{\ell}{m}   & \mathrm{from} &\mathrm{O}_{\hat{\eta}}(2,2+16,\Z{})\;,
\end{array}
\end{equation}
and between $M_* \in\mathrm{GSp}(4,\Z{})$ and 
$\hat\Sigma_* \in \mathrm{O}_{\hat{\eta}}(2,2+16,\Z{})$ for \CP, see table~1 of 
ref.~\cite{Baur:2020yjl} and also 
refs.~\cite{LopesCardoso:1994is,Bailin:1998yt,Malmendier:2014uka,Font:2016odl,Font:2020rsk}. Here, 
the generators $\mathrm{S}$ and $\mathrm{T}$ of the modular group $\SL{2,\Z{}}$ are defined as
\begin{equation}
\mathrm{S} ~=~
\begin{pmatrix}
0&1\\-1&0
\end{pmatrix} \quad\mathrm{and}\quad
\mathrm{T} ~=~
\begin{pmatrix}
1&1\\0&1
\end{pmatrix}\;,
\end{equation}
such that $M_{(\gamma_T,\gamma_U)} \in \mathrm{Sp}(4,\Z{})$ for $\gamma_T\in\SL{2,\Z{}}_T$ and 
$\gamma_U\in\SL{2,\Z{}}_U$. Note that we have changed the definition of the mirror symmetry 
generator $\hat{M}\in\mathrm{O}_{\hat{\eta}}(2,2+16,\Z{})$ as explained in appendix~\ref{app:mirror}.

\section[Stabilizing moduli by Sp(4,Z) orbifolds]{\boldmath Stabilizing moduli by $\mathrm{Sp}(4,\Z{})$ orbifolds\unboldmath}
\label{sec:StabilizedModuliByOrb}

In the following, we consider all inequivalent fixed points of $\mathrm{Sp}(4,\Z{})$ in the Siegel 
upper half-plane $\mathcal{H}_2$, as listed in table~2 of ref.~\cite{Ding:2020zxw}. For each fixed 
point $\tau_\mathrm{f}\in\mathcal{H}_2$, we explicitly construct an orbifold compactification in 
the Narain formulation by specifying a Narain point group $\hat{P}_\mathrm{Narain}$. Then, the 
string moduli are stabilized geometrically by the orbifold action in agreement with the fixed 
point $\tau_\mathrm{f}$. To do so, we focus on the moduli $(T,U,Z)$ of a $D=2$ subsector of a full 
six-dimensional string compactification. In more detail, for each fixed point $\tau_\mathrm{f}$, we 
consider the stabilizer group $\bar{H}:=H/\{\pm\Id_4\}$ from appendix~D of 
ref.~\cite{Ding:2020zxw}, where
\begin{equation}
H := \Big\{\;\gamma = \begin{pmatrix}A&B\\C&D\end{pmatrix}\in \mathrm{Sp}(4,\Z{}) ~\Big|~ \gamma\, \tau_\mathrm{f} ~=~ \tau_\mathrm{f}\;\Big\} \;\;\mathrm{and}\;\; \gamma\, \tau := \left(A\,\tau+B\right) \left(C\,\tau+D\right)^{-1}\;.
\end{equation}
We take the generators of $\bar{H}$ and write them in terms of $\mathrm{Sp}(4,\Z{})$ basis elements 
using the notation of ref.~\cite{Baur:2020yjl}. Then, we apply the dictionary 
eq.~\eqref{eq:SP4ZBasisElements} between $\mathrm{Sp}(4,\Z{})$ and 
$\mathrm{O}_{\hat{\eta}}(2,2+16,\Z{})$ in order to translate the stabilizer group $\bar{H}$ into a 
subgroup $\hat{P}_\mathrm{Narain}$ of $\mathrm{O}_{\hat{\eta}}(2,2+16,\Z{})$. By construction, 
$\hat{P}_\mathrm{Narain}$ maps the Narain lattice in $D=2$ to itself. Consequently, we can utilize 
$\hat{P}_\mathrm{Narain}$ as a Narain point group to define a Narain space group 
$\hat{S}_\mathrm{Narain}$ without roto-translations. It turns out that the moduli $(T,U,Z)$ of the 
resulting (a)symmetric orbifold are fixed by the orbifold action due to 
condition~\eqref{eq:GeneralizedMetricInvariant}. Hence, for each fixed point $\tau_\mathrm{f}$ of 
$\mathrm{Sp}(4,\Z{})$ listed in ref.~\cite{Ding:2020zxw}, we verify that the string moduli 
$(T,U,Z)$ are fixed accordingly, i.e.\
\begin{equation}
\tau_\mathrm{f} ~=~ \begin{pmatrix}\tau_1&\tau_3\\\tau_3&\tau_2\end{pmatrix} ~\in~ \mathcal{H}_2 \quad\Leftrightarrow\quad \begin{pmatrix}U&Z\\Z&T\end{pmatrix} ~\in~ \mathcal{H}_2\;,
\end{equation}
using an appropriate orbifold compactification. In addition, we use the dictionary between 
$\mathrm{Sp}(4,\Z{})$ and $\mathrm{O}_{\hat{\eta}}(2,2+16,\Z{})$ to translate both, the normalizer 
$N(H)$ from ref.~\cite{Ding:2020zxw} and the \CP transformation from ref.~\cite{Ding:2021iqp} into 
$\mathrm{O}_{\hat{\eta}}(2,2+16,\Z{})$. We check explicitly that the resulting transformations are 
outer automorphisms of the corresponding Narain space group. Hence, for each orbifold we identify 
the modular group $\mathcal{G}_\mathrm{modular}$, including a \CP-like transformation, and compute the 
transformations of the unfixed moduli with respect to the modular generators 
$\hat\Sigma \in \mathcal{G}_\mathrm{modular}$.

Finally, let us remark that the Narain point groups $\hat{P}_\mathrm{Narain}$ that we construct 
for each inequivalent fixed point of $\mathrm{Sp}(4,\Z{})$ are the ``maximal'' point groups that 
one can use for the given fixed point. In other words, one can also consider a subgroup of 
$\hat{P}_\mathrm{Narain}$ as the point group of an orbifold. For example, we will encounter a 
$\Z{6}$ Narain point group for a specific fixed point of $\mathrm{Sp}(4,\Z{})$. In this case, also 
the $\Z{3}$ subgroup of $\Z{6}$ can serve as a point group that will geometrically stabilize the 
moduli in the same way as the $\Z{6}$ orbifold.

\subsection{Orbifolds with moduli spaces of dimension 2}

According to ref.~\cite{Ding:2020zxw}, there are two inequivalent subspaces of complex dimension 2 
that are left invariant by subgroups of $\mathrm{Sp}(4,\Z{})$. As we show in the following, they 
can be implemented in string theory by the compactification on $\Z{2}$ orbifolds.

\subsubsection[Symmetric Z2 orbifold]{\boldmath Symmetric $\Z{2}$ orbifold\unboldmath}
\label{sec:Z2Symmetric}

Let us consider the point 
\begin{equation}\label{eq:FPForSymmetricZ2}
\tau_\mathrm{f} ~=~ \begin{pmatrix}\tau_1&0\\0&\tau_2\end{pmatrix} ~\in~ \mathcal{H}_2
\end{equation}
in the Siegel upper half-plane $\mathcal{H}_2$, i.e.\ a complex two-dimensional subspace of 
$\mathcal{H}_2$ given by $\tau_3=0$. In the following, we will explicitly construct a string 
compactification such that the string moduli are fixed accordingly. To do so, we first have to 
consider the stabilizer $\bar{H}$ of $\tau_\mathrm{f}$. For the point eq.~\eqref{eq:FPForSymmetricZ2} 
the stabilizer is $\bar{H} \cong \Z{2}$, generated by
\begin{equation}
h ~:=~ \begin{pmatrix}
 1& 0& 0& 0\\
 0&-1& 0& 0\\
 0& 0& 1& 0\\
 0& 0& 0&-1
\end{pmatrix} ~\in~ \mathrm{Sp}(4,\Z{}) \quad\mathrm{such\ that}\quad h\,\tau_\mathrm{f} ~=~ \tau_\mathrm{f}\;,
\end{equation}
see ref.~\cite[table~2 and appendix~D]{Ding:2020zxw}. Using the notation of 
ref.~\cite{Baur:2020yjl}, this element $h$ is given by the square of a modular $\mathrm{S}$ 
transformation of the modulus $\tau_2$,
\begin{equation}
h ~=~ M_{(\mathrm{S}^2,\Id_2)} ~\in~ \mathrm{Sp}(4,\Z{})\;.
\end{equation}
From the dictionary eq.~\eqref{eq:SP4ZBasisElements} we know that this $\mathrm{Sp}(4,\Z{})$ 
element corresponds to the $\mathrm{O}_{\hat{\eta}}(2,2+16,\Z{})$ element
\begin{equation}\label{eq:SymZ2NarainTwist}
\hat\Theta ~:=~ \left(\hat{K}_\mathrm{S}\right)^2 ~\in~ \mathrm{O}_{\hat{\eta}}(2,2+16,\Z{})\;.
\end{equation}
As a remark, in $\mathrm{O}_{\hat{\eta}}(2,2+16,\Z{})$ we have 
$(\hat{K}_\mathrm{S})^2 = (\hat{C}_\mathrm{S})^2$. So, we could equally write 
$\hat\Theta=(\hat{C}_\mathrm{S})^2$. This Narain twist $\hat\Theta$ defines a $\Z{2}$ Narain point 
group $\hat{P}_\mathrm{Narain}$ of a symmetric $\Z{2}$ orbifold. The moduli $(T,U,Z)$ in the 
generalized metric $\mathcal{H}$ are constrained by the invariance 
condition~\eqref{eq:GeneralizedMetricInvariant} under $\hat\Theta$. As a result, we find 
$a_1 = a_2 = 0$. Hence, the Wilson line modulus $Z$ has to vanish, $Z=0$. On the other hand, the 
moduli
\begin{equation}
T ~=~ \frac{1}{\alpha'}\left(B_{12}+\I\,\sqrt{\mathrm{det}\,G}\right) \quad\mathrm{and}\quad
U ~=~ \frac{1}{G_{11}}\left(G_{12}+\I\,\sqrt{\mathrm{det}\,G}\right)
\end{equation}
are unconstrained, as it has been known from ref.~\cite{Baur:2020jwc,Baur:2021mtl}, for example. In 
other words, the Narain lattice is mapped to itself under the Narain twist $\hat\Theta$ 
eq.~\eqref{eq:SymZ2NarainTwist} only if the Wilson line modulus is trivial $Z=0$, while the 
K\"ahler modulus $T$ and the complex structure modulus $U$ can vary freely. One says that $Z$ has 
been stabilized geometrically by the orbifold action. This is in agreement with $\tau_\mathrm{f}$ 
under the identification $\tau_1=U$, $\tau_2=T$ and $\tau_3=Z=0$. In other words, the complex 
two-dimensional subspace with $\tau_3=0$ described in ref.~\cite{Ding:2020zxw} can be constructed 
in string theory by the compactification on a symmetric $\Z{2}$ orbifold with vanishing Wilson line.

Next, we translate the normalizer
\begin{equation}
N(H) ~=~ \Big\langle\;M_{(\mathrm{S},\Id_2)}\;,\;M_{(\mathrm{T},\Id_2)}\;,\;M_{(\Id_2,\mathrm{S})}\;,\;M_{(\Id_2,\mathrm{T})}\;,\; M_{\times}\;\Big\rangle ~\subset~ \mathrm{Sp}(4,\Z{})
\end{equation}
from ref.~\cite{Ding:2020zxw} into $\mathrm{O}_{\hat{\eta}}(2,2+16,\Z{})$ and check that each 
generator is an outer automorphism of the Narain space group, see eq.~\eqref{eq:GModularOfOrbifold}. 
In this way, we identify the modular group (without \CP)
\begin{equation}
\mathcal{G}_\mathrm{modular} ~=~ \Big\langle\;\hat{K}_\mathrm{S}\;,\;\hat{K}_\mathrm{T}\;,\;\hat{C}_\mathrm{S}\;,\;\hat{C}_\mathrm{T}\;,\;\hat{M}\;\Big\rangle ~\cong~ \frac{\SL{2,\Z{}}_T \times \SL{2,\Z{}}_U}{\Z{2}} \rtimes \Z{2}^{\hat{M}}
\end{equation}
of the symmetric $\Z{2}$ orbifold, where the $\Z{2}$ quotient ensures the relation 
$(\hat{K}_\mathrm{S})^2 = (\hat{C}_\mathrm{S})^2$, see section 3 of ref.~\cite{Baur:2019iai}. Then, 
we use the generalized metric~\eqref{eq:TrafoOfModuli} in order to identify the transformation of 
the moduli $(T,U)$ and we obtain the transformations~\eqref{eq:ModuliTrafoKS}-\eqref{eq:ModuliTrafoMirror} 
with $Z=0$,
while $Z=0$ is invariant under all of these transformations. As a remark, for the symmetric $\Z{2}$ 
orbifold, the finite modular group $\mathcal{G}_\mathrm{fmg}$ is known explicitly from the 
transformation matrices $\rho_{\rep{r}}(\hat\Sigma)$ of twisted strings with respect to modular 
transformations $\hat\Sigma$. It is given by~\cite{Baur:2020jwc,Baur:2021mtl}
\begin{equation}
\mathcal{G}_\mathrm{fmg} ~=~ \left(S_3^T \times S_3^U\right) \rtimes \Z{4}^{\hat{M}} ~\cong~ [144,115]\;,
\end{equation}
where the $\Z{2}^{\hat{M}}$ mirror symmetry of the moduli acts as a $\Z{4}^{\hat{M}}$ symmetry on 
the twisted strings of the $\Z{2}$ orbifold. Moreover, the four twisted strings localized at the 
four orbifold fixed points transform as $\rep{r}=\rep{4}_1$ of the finite modular group $[144,115]$.

Next, we consider table~1 of ref.~\cite{Ding:2021iqp} and translate their \CP transformation into 
$\mathrm{O}_{\hat{\eta}}(2,2+16,\Z{})$. We obtain
\begin{equation}
\hat\CP ~=~ \hat{\Sigma}_*\;.
\end{equation}
One can verify easily that $\hat\CP$ is an outer automorphism of the Narain space group, i.e.
\begin{equation}
\hat\CP\,\hat\Theta\,\hat\CP^{-1} ~=~ \hat\Theta^{-1}\;.
\end{equation}
Furthermore, we use the generalized metric~\eqref{eq:TrafoOfModuli} and confirm the transformation 
of the string moduli $(T,U)$,
\begin{equation}
T ~\xmapsto{\hat\CP}~ -\bar{T}\;, \qquad U ~\xmapsto{\hat\CP}~ -\bar{U}\;,
\end{equation}
while $Z=0$ is invariant under $\hat\CP$, see also ref.~\cite{Novichkov:2019sqv}.

\newpage
\subsubsection[Asymmetric Z2 orbifold]{\boldmath Asymmetric $\Z{2}$ orbifold\unboldmath}
\label{sec:Z2Asymmetric}

The second fixed point in the list of ref.~\cite{Ding:2020zxw} is given by
\begin{equation}
\tau_\mathrm{f} ~=~ \begin{pmatrix}\tau_1&\tau_3\\\tau_3&\tau_1\end{pmatrix} ~\in~ \mathcal{H}_2\;.
\end{equation}
In this case, the stabilizer $\bar{H} \cong \Z{2}$ is generated by the mirror element
\begin{equation}\label{eq:h1beforeBasisChange}
h ~:=~ \begin{pmatrix}
 0& 1& 0& 0\\
 1& 0& 0& 0\\
 0& 0& 0& 1\\
 0& 0& 1& 0
\end{pmatrix} ~=~ M_{\times} ~\in~ \mathrm{Sp}(4,\Z{})\,, \quad\text{such that}\quad h\,\tau_\mathrm{f} ~=~ \tau_\mathrm{f}
\end{equation}
in the notation of ref.~\cite{Baur:2020yjl}. Since $\hat{M}\in\mathrm{O}_{\hat{\eta}}(2,2+16,\Z{})$ 
denotes the corresponding mirror element in the string constructing, we choose a Narain twist
\begin{equation}
\hat\Theta ~:=~ \hat{M}\;,
\end{equation}
see eq.~\eqref{eq:NewMirrorMatrix} in the appendix. Consequently, we construct an asymmetric 
$\Z{2}$ orbifold with mirror symmetry $\hat{M}$ as Narain twist. Then, 
eq.~\eqref{eq:GeneralizedMetricInvariant} yields 
\begin{equation}\label{eq:G11anda1}
G_{11} ~=~ \alpha'(1-a_1^2) \quad\mathrm{and}\quad B_{12}~=~a_1\, a_2\, \alpha' + G_{12}\;.
\end{equation}
In this case, one can check that the expectations for the corresponding fixed point 
$\tau_\mathrm{f}$ are fulfilled,
\begin{equation}
T ~=~ U\;, \quad\mathrm{where}\quad U ~=~ \frac{1}{\alpha'\left(1-a_1^2\right)}\,\left(G_{12} + \I\, \sqrt{-G_{12}^2+\alpha'\,G_{22}\,\left(1-a_1^2\right)}\right)\;,
\end{equation}
and $Z = -a_2+U a_1$. Hence, the two-dimensional subspace with $\tau_1=\tau_2$ and an unconstrained 
$\tau_3$, as described in the bottom-up construction of ref.~\cite{Ding:2020zxw}, can be obtained 
in string theory from the compactification on an asymmetric $\Z{2}$ orbifold using mirror symmetry 
as orbifold twist. Note that a physical torus must satisfy $G_{11} > 0$. Thus, 
eq.~\eqref{eq:G11anda1} constrains the Wilson line to $a_1^2 < 1$. Let us remark that the 
Narain twist $\hat\Theta$ given by the mirror transformation $\hat{M}$ also acts on the 16 gauge 
degrees of freedom of the heterotic string. This will induce some gauge symmetry breaking, 
cf.~refs.~\cite{Ibanez:1987xa,Forste:2005rs}.

Next, we consider the normalizer 
\begin{equation}
N(H) ~=~ \Big\langle\;M_{(\mathrm{S},\mathrm{S})}\;,\;M_{(\mathrm{T},\mathrm{T})}\;,\;M_{(\mathrm{S}^2,\Id_2)}\;,\;\M{-1}{0}\;\Big\rangle ~\subset~ \mathrm{Sp}(4,\Z{})\;,
\end{equation}
given in ref.~\cite{Ding:2020zxw}. We use our dictionary eq.~\eqref{eq:SP4ZBasisElements} and 
obtain the group 
\begin{equation}\label{eq:GModularForAsymmetricZ2}
\mathcal{G}_\mathrm{modular} ~=~ \Big\langle\;\hat{K}_\mathrm{S}\,\hat{C}_\mathrm{S}\;,\;\hat{K}_\mathrm{T}\,\hat{C}_\mathrm{T}\;,\;\left(\hat{K}_\mathrm{S}\right)^2\;,\;\W{-1}{0}\;\Big\rangle\;.
\end{equation}
We verify that this group gives rise to the rotational outer automorphisms of the Narain space 
group. Hence, $\mathcal{G}_\mathrm{modular}$ is the modular group of the asymmetric $\Z{2}$ orbifold. Using 
eq.~\eqref{eq:TrafoOfModuli} we find that the two independent moduli $(T,Z)$ transform as
\begin{subequations}
\begin{align}
T &~\xmapsto{\hat{K}_\mathrm{S}\,\hat{C}_\mathrm{S}}~ -\frac{T}{T^2-Z^2}\;, &Z &~\xmapsto{\hat{K}_\mathrm{S}\,\hat{C}_\mathrm{S}}~ \frac{Z}{T^2-Z^2}\;,\\
T &~\xmapsto{\hat{K}_\mathrm{T}\,\hat{C}_\mathrm{T}}~ T + 1\;,              &Z &~\xmapsto{\hat{K}_\mathrm{T}\,\hat{C}_\mathrm{T}}~ Z\;,\\
T &~\xmapsto{\left(\hat{K}_\mathrm{S}\right)^2}~      T\;,                  &Z &~\xmapsto{\left(\hat{K}_\mathrm{S}\right)^2}~ -Z\;,\\
T &~\xmapsto{\W{-1}{0}}~      T\;,                                          &Z &~\xmapsto{\W{-1}{0}}~ Z+1\;,
\end{align}
\end{subequations}
under the generators of the modular symmetry $\mathcal{G}_\mathrm{modular}$ of this asymmetric orbifold.

Finally, we consider the \CP transformation for the fixed point $\tau_\mathrm{f}$ with 
$\tau_1=\tau_2$ in table~1 of ref.~\cite{Ding:2021iqp}. Using our dictionary to 
$\mathrm{O}_{\hat{\eta}}(2,2+16,\Z{})$, this transformation corresponds to
\begin{equation}\label{eq:CPOfAsymZ2}
\hat\CP ~=~ \hat{\Sigma}_*\ \qquad\mathrm{with}\qquad \hat\CP\,\hat\Theta\,\hat\CP^{-1} ~=~ \hat\Theta^{-1}\;.
\end{equation}
Hence, $\hat\CP$ is an outer automorphism of $\hat{S}_\mathrm{Narain}$. From the generalized 
metric~\eqref{eq:TrafoOfModuli} we compute the transformation of the string moduli $(T,Z)$, 
resulting in
\begin{equation}
T ~\xmapsto{\hat\CP}~ -\bar{T}\;, \qquad Z ~\xmapsto{\hat\CP}~ -\bar{Z}\;,
\end{equation}
for this asymmetric $\Z{2}$ orbifold.

\subsubsection[Symmetric Z2 orbifold with discrete Wilson line]{\boldmath Symmetric $\Z{2}$ orbifold with discrete Wilson line\unboldmath}
\label{sec:Z2SymmetricWL}

Let us briefly show that the asymmetric \Z{2} orbifold constructed in section~\ref{sec:Z2Asymmetric} 
is equivalent to a symmetric \Z{2} orbifold with nontrivial discrete Wilson line~\cite{Ibanez:1986tp,Erler:1991ju}. 
According to appendix D.1 of ref.~\cite{Ding:2020zxw}, the fixed point $\tau_\mathrm{f}$ (with 
$\tau_1=\tau_2$) is equivalent to a fixed point $\tau'_\mathrm{f}$ using a $\mathrm{Sp}(4,\Z{})$ 
transformation
\begin{equation}\label{eq:Z2BasisChangeinSP4Z}
b ~:=~ \begin{pmatrix}
 1& 0& 0& 0\\
 1& 0& 0&-1\\
 0&-1& 1& 0\\
 0& 1& 0& 0
\end{pmatrix} ~=~ M_{(\mathrm{T}^{-1},\Id_2)}\,\M{0}{1}\,M_{(\mathrm{T}\mathrm{S}^3,\Id_2)} ~\in~ \mathrm{Sp}(4,\Z{})\;,
\end{equation}
such that
\begin{equation}\
b\,\tau'_\mathrm{f} ~=~ \tau_\mathrm{f}\;, \quad\mathrm{where}\quad \tau_\mathrm{f} ~:=~ \begin{pmatrix}\tau_1&\tau_3\\\tau_3&\tau_1\end{pmatrix} \quad\mathrm{and}\quad \tau'_\mathrm{f} ~=~ \begin{pmatrix}\tau'_1&\nicefrac{1}{2}\\\nicefrac{1}{2}&\tau'_2\end{pmatrix}\;.
\end{equation}
Here, $\tau'_1 := \frac{1}{2}\left(\tau_3+\tau_1\right)$ and $\tau'_2 := \frac{1}{2\left(\tau_3-\tau_1\right)}$. 
Since $\tau_\mathrm{f}$ is a fixed point of $h=M_{\times}\in\mathrm{Sp}(4,\Z{})$, see 
eq.~\eqref{eq:h1beforeBasisChange}, we find
\begin{equation}
\tau_\mathrm{f} ~=~ h\, \tau_\mathrm{f} \quad\Leftrightarrow\quad b\,\tau'_\mathrm{f} ~=~ h\, b\,\tau'_\mathrm{f} \quad\Leftrightarrow\quad \tau'_\mathrm{f} ~=~ \left(b^{-1}h\, b\right)\,\tau'_\mathrm{f}\;,
\end{equation}
such that $\tau'_\mathrm{f}$ is a fixed point of $\left(b^{-1}h\, b\right)\in\mathrm{Sp}(4,\Z{})$. 
Next, we map
\begin{equation}\label{eq:Sp4ZZ2TrafoAfterBasisChange}
h'~:=~\left(b^{-1}h\, b\right) ~=~ 
\begin{pmatrix}
 1& 0& 0&-1\\
 0&-1& 1& 0\\
 0& 0& 1& 0\\
 0& 0& 0&-1
\end{pmatrix} ~=~
M_{(\mathrm{S}^2,\Id_2)}\,\M{1}{0}
\end{equation}
to the corresponding Narain twist $\hat\Theta'\in\mathrm{O}_{\hat{\eta}}(2,2+16,\Z{})$,
\begin{equation}\label{eq:AsymZ2ThetaPrime}
\hat\Theta' ~:=~ \hat{B}^{-1}\,\hat\Theta\,\hat{B}\;.
\end{equation}
Here, $\hat{B}\in\mathrm{O}_{\hat{\eta}}(2,2+16,\Z{})$ corresponds to $b\in\mathrm{Sp}(4,\Z{})$ 
from eq.~\eqref{eq:Z2BasisChangeinSP4Z}, i.e.\
\begin{equation}
\hat{B} ~:=~ \left(\hat{K}_\mathrm{T}\right)^{-1}\,\W{0}{1}\,\hat{K}_\mathrm{T}\,\left(\hat{K}_\mathrm{S}\right)^3\;.
\end{equation}
Since $\hat\Theta$ and $\hat\Theta'$ are related in eq.~\eqref{eq:AsymZ2ThetaPrime} by conjugation 
with $\hat{B}\in\mathrm{O}_{\hat{\eta}}(2,2+16,\Z{})$, the two Narain point groups belong to the 
same Narain \Z{}-class~\cite{GrootNibbelink:2017usl}: they describe the same physics but in 
different duality frames. Now, we use $\hat\Theta:=\hat{M}$ and simplify the Narain twist 
$\hat\Theta'$ from eq.~\eqref{eq:AsymZ2ThetaPrime}, 
\begin{equation}\label{eq:Z2NarainTwistThetaPrime}
\hat\Theta' ~=~ \left(\hat{K}_\mathrm{S}\right)^2\,\W{1}{0} ~=~ \left(\hat{C}_\mathrm{S}\right)^2\,\W{1}{0}\;,
\end{equation}
in agreement with eq.~\eqref{eq:Sp4ZZ2TrafoAfterBasisChange}. As a result, we have shown that the 
asymmetric $\Z{2}$ orbifold with a Narain twist $\hat\Theta$ given by mirror symmetry is equivalent 
to a symmetric $\Z{2}$ orbifold with Narain twist $\hat\Theta'$. Then, the new Narain twist 
$\hat\Theta'$ constrains the generalized metric eq.~\eqref{eq:GeneralizedMetricInvariant} such that 
the Wilson lines have to be fixed,
\begin{equation}
a_1 ~=~ 0 \quad\mathrm{and}\quad a_2 ~=~ \nicefrac{-1}{2}\;.
\end{equation}
Consequently, the moduli $(T,U,Z)$ of the symmetric $\Z{2}$ orbifold read
\begin{equation}
Z ~=~ \nicefrac{1}{2} \quad\mathrm{and}\quad T,\ U \mathrm{\ unconstrained}\;,
\end{equation}
as expected for $\tau'_\mathrm{f}$. The modular group of this orbifold can be obtained from 
eq.~\eqref{eq:GModularForAsymmetricZ2} using the basis change $\hat{B}$,
\begin{equation}
\mathcal{G}_\mathrm{modular} ~=~ \Big\langle\;\hat{M}_1\;,\;\hat{M}_2\;,\;\hat{M}_3\;,\;\hat{M}_4\;\Big\rangle\;,
\end{equation}
where we have defined
\begin{subequations}
\begin{align}
\hat{M}_1 & ~:=~ \hat{B}^{-1}\,\hat{K}_\mathrm{S}\,\hat{C}_\mathrm{S}\,\hat{B}\;, & \hat{M}_2 & ~:=~ \hat{B}^{-1}\,\hat{K}_\mathrm{T}\,\hat{C}_\mathrm{T}\,\hat{B}\;,\\
\hat{M}_3 & ~:=~ \hat{B}^{-1}\,\left(\hat{K}_\mathrm{S}\right)^2\,\hat{B}\;,      & \hat{M}_4 & ~:=~ \hat{B}^{-1}\,\W{-1}{0}\,\hat{B}\;.
\end{align}
\end{subequations}
Note that these transformations satisfy the relations
\begin{equation}
\left(\hat{M}_1\right)^2 ~=~ \left(\hat{M}_3\right)^2 ~=~ \left(\hat{M}_1\,\hat{M}_2\right)^3 ~=~ \left(\hat{M}_1\,\hat{M}_4\right)^6 ~=~ \left(\hat{M}_3\,\hat{M}_4\right)^2 ~=~ \Id_{20}\;.
\end{equation}
Then, we use eq.~\eqref{eq:TrafoOfModuli} to compute the modular transformations of the moduli 
$(T,U)$,
\begin{subequations}\label{eq:ModGroupOnModuliNewBasis}
\begin{align}
T &~\xmapsto{\hat{M}_1}~ -\frac{1}{4T}\;,    &U &~\xmapsto{\hat{M}_1}~ -\frac{1}{4U}\;,\\
T &~\xmapsto{\hat{M}_2}~ -\frac{T}{2\,T-1}\;,&U &~\xmapsto{\hat{M}_2}~ U + \frac{1}{2}\;,\\
T &~\xmapsto{\hat{M}_3}~ -\frac{1}{4U}\;,    &U &~\xmapsto{\hat{M}_3}~ -\frac{1}{4T}\;,\\
T &~\xmapsto{\hat{M}_4}~  \frac{T}{2\,T+1}\;,&U &~\xmapsto{\hat{M}_4}~ U + \frac{1}{2}\;,
\end{align}
\end{subequations}
while $Z = \nicefrac{1}{2}$ is invariant under all of these transformations. Using the relations
\begin{equation}
\hat{M}_1\,\hat{M}_2\,\left(\hat{M}_4\right)^{-1}\hat{M}_1 ~=~ \hat{K}_\mathrm{T}\quad,\quad
\hat{M}_2\,\hat{M}_4                                       ~=~ \hat{C}_\mathrm{T}\quad\mathrm{and}\quad
\hat{M}_1\,\hat{M}_3                                       ~=~ \hat{M}\,\hat\Theta'\;,
\end{equation}
we observe that the transformations
\begin{subequations}
\begin{align}
T &~\xmapsto{~\hat{K}_\mathrm{T}~}~ T+1\;, &U &~\xmapsto{~\hat{K}_\mathrm{T}~}~ U\;,\\
T &~\xmapsto{~\hat{C}_\mathrm{T}~}~ T\;,   &U &~\xmapsto{~\hat{C}_\mathrm{T}~}~ U+1\;,\\
T &~\xmapsto{\hat{M}\,\hat\Theta'}~ U\;,   &U &~\xmapsto{\hat{M}\,\hat\Theta'}~ T\;,
\end{align}
\end{subequations}
follow from eqs.~\eqref{eq:ModGroupOnModuliNewBasis}. Thus, there exists an alternative set of 
generators of $\mathcal{G}_\mathrm{modular}$ given by $\hat{K}_\mathrm{T}$, $\hat{M}\,\hat\Theta'$, 
$\hat{M}_1$ and $\hat{M}_2$.

Finally, we analyze \CP for the fixed point $\tau'_\mathrm{f}$ with $\tau'_3=\nicefrac{1}{2}$. 
Using the basis change $\hat{B}$ and $\hat\CP = \hat{\Sigma}_*$ from eq.~\eqref{eq:CPOfAsymZ2}, we 
obtain
\begin{equation}\label{eq:CPZ2withWL}
\hat\CP' ~=~ \hat{B}^{-1}\,\hat{\Sigma}_*\,\hat{B} ~=~ \left(\hat{K}_\mathrm{S}\right)^2\hat{\Sigma}_* \qquad\mathrm{with}\qquad \hat\CP'\,\hat\Theta'\,\hat\CP'^{-1} ~=~ \hat\Theta'^{-1}\;.
\end{equation}
Thus, $\hat\CP'$ is an outer automorphism of the Narain space group of the symmetric $\Z{2}$ 
orbifold with discrete Wilson line. The generalized metric~\eqref{eq:TrafoOfModuli} transforms 
under $\hat\CP'$ such that we find
\begin{equation}
T ~\xmapsto{\hat\CP'}~ -\bar{T}\;, \qquad U ~\xmapsto{\hat\CP'}~ -\bar{U}\;,
\end{equation}
while $Z=\nicefrac{1}{2}$ is invariant under the transformation with $\hat\CP'$.

\subsection{Orbifolds with moduli spaces of dimension 1}

Following ref.~\cite{Ding:2020zxw}, there are five inequivalent subspaces of complex dimension 1 
that are left invariant by subgroups of $\mathrm{Sp}(4,\Z{})$. In the following, we implement them 
explicitly in string theory by the compactification on various symmetric and asymmetric orbifolds.

\subsubsection[Symmetric Z4 orbifold]{\boldmath Symmetric $\Z{4}$ orbifold\unboldmath}
\label{sec:Z4Symmetric}

Consider the fixed point
\begin{equation}
\tau_\mathrm{f} ~=~ \begin{pmatrix}\I&0\\0&\tau_2\end{pmatrix} ~\in~ \mathcal{H}_2\;.
\end{equation}
In this case, the stabilizer $\bar{H} \cong \Z{4}$ is generated by
\begin{equation}
h ~:=~ \begin{pmatrix}
 0& 0& 1& 0\\
 0& 1& 0& 0\\
-1& 0& 0& 0\\
 0& 0& 0& 1
\end{pmatrix} ~=~ M_{(\Id_2,\mathrm{S})} ~\in~ \mathrm{Sp}(4,\Z{})\;,
\end{equation}
using the notation of ref.~\cite{Baur:2020yjl}. This $\mathrm{Sp}(4,\Z{})$ element corresponds in 
string theory to a modular $\mathrm{S}$ transformation of the complex structure modulus. Thus, we 
choose a Narain twist
\begin{equation}
\hat\Theta ~:=~ \hat{C}_\mathrm{S} ~\in~ \mathrm{O}_{\hat{\eta}}(2,2+16,\Z{})\;.
\end{equation}
As a result, we construct a symmetric $\Z{4}$ orbifold with Narain twist 
$\hat\Theta = \hat{C}_\mathrm{S}$. In order to satisfy eq.~\eqref{eq:GeneralizedMetricInvariant} 
we have to set 
\begin{equation}
G_{11} ~=~ G_{22}\quad,\quad G_{12} ~=~ 0\quad\mathrm{and}\quad a_1 ~=~ a_2 ~=~ 0\;.
\end{equation}
Hence, we have to fix the Wilson lines to zero $Z=0$, the complex structure modulus to $U=\I$, 
while the K\"ahler modulus $T$ remains unconstrained, 
\begin{equation}
T ~=~ \frac{1}{\alpha'}\,\left(B_{12}+\I\,G_{11}\right)\;,
\end{equation}
as expected for the values of $\tau_\mathrm{f}$ in this case. 

The normalizer is given by
\begin{equation}
N(H) ~=~ \Big\langle \;M_{(\mathrm{S},\Id_2)}\;,\;M_{(\mathrm{T},\Id_2)}\;,\;M_{(\Id_2,\mathrm{S})}\;\Big\rangle ~\subset~ \mathrm{Sp}(4,\Z{})\;,
\end{equation}
see ref.~\cite{Ding:2020zxw}. Compared to ref.~\cite{Ding:2020zxw}, we use $M_{(\mathrm{S},\Id_2)}$ 
and $M_{(\Id_2,\mathrm{S})}$ as generators of $N(H)$ instead of $M_{(\mathrm{S}^3,\Id_2)}$ and 
$M_{(\Id_2,\mathrm{S}^3)}$. The corresponding group in $\mathrm{O}_{\hat{\eta}}(2,2+16,\Z{})$ reads
\begin{equation}
\mathcal{G}_\mathrm{modular} ~=~ \Big\langle\;\hat{K}_\mathrm{S}\;,\;\hat{K}_\mathrm{T}\;,\;\,\hat{C}_\mathrm{S}\;\Big\rangle ~\cong~ \left(\SL{2,\Z{}}_T ~\times~ \Z{4}^R\right)/~\Z{2}\;,
\end{equation}
which is the modular group of the symmetric $\Z{4}$ orbifold. Note that $\Z{4}^R$ allows for a 
geometrical interpretation as a $\nicefrac{\pi}{2}$ sublattice rotation, assuming that this $D=2$ 
orbifold is a subsector of a full six-dimensional string compactification, see for example 
section~3 of ref.~\cite{Nilles:2020gvu}. Consequently, $\Z{4}^R$ is an $R$-symmetry, as our 
notation explicitly indicates. Note that the order of this geometrical $\Z{4}$ sublattice rotation 
will generically be larger than four due to fractional $R$-charges of twisted strings, cf.\ 
refs.~\cite{Kobayashi:2004ya,Bizet:2013gf,Nilles:2013lda,Bizet:2013wha,Nilles:2017heg,Nilles:2020tdp,Nilles:2020gvu}.
Using the generalized metric~\eqref{eq:TrafoOfModuli}, we confirm that the K\"ahler modulus $T$ 
transforms under the generators of the modular group $\mathcal{G}_\mathrm{modular}$ as expected, i.e.\
\begin{equation}
T ~\xmapsto{\hat{K}_\mathrm{S}}~ -\frac{1}{T}\quad,\quad
T ~\xmapsto{\hat{K}_\mathrm{T}}~ T+1\quad\mathrm{and}\quad
T ~\xmapsto{\hat{C}_\mathrm{S}}~ T\;.
\end{equation}

Finally, we consider the \CP transformation for the fixed point $\tau_\mathrm{f}$ with 
$\tau_1=\I$ and $\tau_3=0$ given in table~1 of ref.~\cite{Ding:2021iqp}. Using our dictionary 
eq.~\eqref{eq:SP4ZBasisElements} from $\mathrm{GSp}(4,\Z{})$ to $\mathrm{O}_{\hat{\eta}}(2,2+16,\Z{})$, 
the corresponding \CP transformation of this symmetric $\Z{4}$ orbifold is given by
\begin{equation}
\hat\CP ~=~ \hat{\Sigma}_* \qquad\mathrm{with}\qquad \hat\CP\,\hat\Theta\,\hat\CP^{-1} ~=~ \hat\Theta^{-1}\;.
\end{equation}
Thus, it is an outer automorphism of $\hat{S}_\mathrm{Narain}$. We apply $\hat\CP$ to the 
generalized metric~\eqref{eq:TrafoOfModuli} and find that $U=\I$ and $Z=0$ are invariant, while the 
K\"ahler modulus transforms as
\begin{equation}
T ~\xmapsto{\hat\CP}~ -\bar{T}\;.
\end{equation}

\newpage

\subsubsection[Symmetric Z6 orbifold]{\boldmath Symmetric $\Z{6}$ orbifold\unboldmath}
\label{sec:Z6Symmetric}

The next fixed point in the list of ref.~\cite{Ding:2020zxw} is given by
\begin{equation}
\tau_\mathrm{f} ~=~ \begin{pmatrix}\omega&0\\0&\tau_1\end{pmatrix} ~\in~ \mathcal{H}_2\;,
\end{equation}
where $\omega :=\exp(\nicefrac{2\pi\I}{3})$. In this case, the stabilizer $\bar{H} \cong \Z{6}$ is 
generated by one element that we can write in the notation of ref.~\cite{Baur:2020yjl} as
\begin{equation}
h ~:=~ \begin{pmatrix}
 1& 0& 1& 0\\
 0& 1& 0& 0\\
-1& 0& 0& 0\\
 0& 0& 0& 1
\end{pmatrix} ~=~ M_{(\Id_2,\mathrm{S}^3\mathrm{T}\mathrm{S}\mathrm{T})} ~\in~ \mathrm{Sp}(4,\Z{})\;.
\end{equation}
Hence, we choose a Narain twist
\begin{equation}\label{eq:SymZ6NarainTwist}
\hat\Theta ~:=~ \left(\hat{C}_\mathrm{S}\right)^3\hat{C}_\mathrm{T}\,\hat{C}_\mathrm{S}\,\hat{C}_\mathrm{T} ~\in~ \mathrm{O}_{\hat{\eta}}(2,2+16,\Z{})\;.
\end{equation}
This Narain twist defines a symmetric $\Z{6}$ orbifold. In order to satisfy 
eq.~\eqref{eq:GeneralizedMetricInvariant}, we have to fix the string geometry as follows 
\begin{equation}
G_{11} ~=~ -2\,G_{12}\quad,\quad G_{22} ~=~ G_{11}\quad\mathrm{and}\quad a_1 ~=~ a_2 ~=~ 0\;.
\end{equation}
This results in
\begin{equation}
U~=~\omega \quad\mathrm{and}\quad Z~=~0\;,
\end{equation}
while the K\"ahler modulus $T$ is unconstrained, as expected by comparing to the value of 
$\tau_\mathrm{f}$ in the corresponding bottom-up construction of ref.~\cite{Ding:2020zxw}.

In this case, the normalizer reads
\begin{equation}
N(H) ~=~ \Big\langle\;M_{(\mathrm{S},\Id_2)}\;,\;M_{(\mathrm{T},\Id_2)}\;,\;M_{(\Id_2,\mathrm{S}^3\mathrm{T})}\;\Big\rangle ~\subset~ \mathrm{Sp}(4,\Z{})\;,
\end{equation}
see ref.~\cite{Ding:2020zxw}. Compared to ref.~\cite{Ding:2020zxw}, we use 
$M_{(\mathrm{S},\Id_2)}$ as generator of $N(H)$ instead of $M_{(\mathrm{S}^3,\Id_2)}$. The 
corresponding group in $\mathrm{O}_{\hat{\eta}}(2,2+16,\Z{})$ gives the modular group 
\begin{equation}
\mathcal{G}_\mathrm{modular} ~=~ \Big\langle\;\hat{K}_\mathrm{S}\;,\;\hat{K}_\mathrm{T}\;,\;\left(\hat{C}_\mathrm{S}\right)^3\hat{C}_\mathrm{T}\;\Big\rangle ~\cong~ \left(\SL{2,\Z{}}_T ~\times~ \Z{6}^R\right)/~\Z{2}
\end{equation}
of the symmetric $\Z{6}$ orbifold, where $(\hat{C}_\mathrm{S})^3\hat{C}_\mathrm{T}$ generates a 
geometrical $\Z{6}$ rotation (and $(\hat{C}_\mathrm{S})^3\hat{C}_\mathrm{T} = \hat\Theta^{-1}$). 
As a sublattice rotation of a six-dimensional orbifold compactification this gives rise to an 
$R$-symmetry, whose order will generically be larger than six due to fractional $R$-charges of 
twisted strings~\cite{Kobayashi:2004ya,Bizet:2013gf,Nilles:2013lda,Bizet:2013wha,Nilles:2017heg,Nilles:2020tdp,Nilles:2020gvu}.
Using the generalized metric~\eqref{eq:TrafoOfModuli}, we compute the transformations of the 
K\"ahler modulus $T$ under the generators of the modular group $\mathcal{G}_\mathrm{modular}$ and 
obtain the expected results, i.e.\
\begin{equation}
T ~\xmapsto{\hat{K}_\mathrm{S}}~ -\frac{1}{T}\quad,\quad
T ~\xmapsto{\hat{K}_\mathrm{T}}~ T+1\quad\mathrm{and}\quad
T ~\xmapsto{\left(\hat{C}_\mathrm{S}\right)^3\hat{C}_\mathrm{T}}~ T\;.
\end{equation}

Finally, we consider the \CP transformation from table~1 of ref.~\cite{Ding:2021iqp} that leaves 
the fixed point $\tau_\mathrm{f}$ with $\tau_1=\omega$ and $\tau_3=0$ invariant. Using our 
dictionary eq.~\eqref{eq:SP4ZBasisElements} from $\mathrm{GSp}(4,\Z{})$ to 
$\mathrm{O}_{\hat{\eta}}(2,2+16,\Z{})$, the corresponding \CP transformation is given by
\begin{equation}
\hat\CP ~=~ \left(\hat{C}_\mathrm{T}\right)^{-1}\hat{\Sigma}_* \qquad\mathrm{with}\qquad \hat\CP\,\hat\Theta\,\hat\CP^{-1} ~=~ \hat\Theta^{-1}\;.
\end{equation}
Hence, $\hat\CP$ is an outer automorphism of the $\Z{6}$ Narain space group. Using 
eq.~\eqref{eq:TrafoOfModuli} for the \CP-like transformation $\hat\CP$, we find that $U=\omega$ and 
$Z=0$ are invariant, while the K\"ahler modulus $T$ transforms as
\begin{equation}
T ~\xmapsto{\hat\CP}~ -\bar{T}\;.
\end{equation}

Let us briefly remark that one can define a symmetric $\Z{3}$ orbifold by taking just the $\Z{3}$ 
subgroup of the stabilizer $\bar{H} \cong \Z{6}$ as Narain point group, i.e.\ consider a $\Z{3}$ 
Narain point group $\hat{P}_\mathrm{Narain}$ generated by the Narain twist $\hat\Theta^2$, where 
$\hat\Theta$ is given in eq.~\eqref{eq:SymZ6NarainTwist}. This also fixes $U=\omega$ and $Z=0$, 
yields the same modular group $\mathcal{G}_\mathrm{modular}$ and the same \CP transformation.

\subsubsection[Asymmetric Z2 x Z2 orbifold]{\boldmath Asymmetric $\Z{2}\times\Z{2}$ orbifold\unboldmath}
\label{sec:Z2xZ2Asymmetric}

For the fixed point
\begin{equation}
\tau_\mathrm{f} ~=~ \begin{pmatrix}\tau_1&0\\0&\tau_1\end{pmatrix} ~\in~ \mathcal{H}_2\;,
\end{equation}
the stabilizer is given by $\bar{H} \cong D_8/\{\pm\Id_4\}$, where $D_8$ is generated by two 
$\mathrm{Sp}(4,\Z{})$ elements,
\begin{equation}
h_1 ~:=~ \begin{pmatrix}
 0& 1& 0& 0\\
 1& 0& 0& 0\\
 0& 0& 0& 1\\
 0& 0& 1& 0
\end{pmatrix} ~=~ M_{\times} \quad\mathrm{and}\quad
h_2 ~:=~ \begin{pmatrix}
 0&-1& 0& 0\\
 1& 0& 0& 0\\
 0& 0& 0&-1\\
 0& 0& 1& 0
\end{pmatrix} ~=~ M_{\times}\, M_{(\mathrm{S}^2,\Id_2)}
\end{equation}
Consequently, we choose two associated Narain twists $\hat\Theta_1, \hat\Theta_2\in\mathrm{O}_{\hat{\eta}}(2,2+16,\Z{})$ as
\begin{equation}
\hat\Theta_1 ~:=~ \hat{M} \quad\mathrm{and}\quad \hat\Theta_2 ~:=~ \hat{M}\,\left(\hat{K}_\mathrm{S}\right)^2\;.
\end{equation}
In $\mathrm{O}_{\hat{\eta}}(2,2+16,\Z{})$, the two Narain twists $\hat\Theta_1$ and $\hat\Theta_2$ 
generate a $\Z{2}\times\Z{2}$ Narain point group $\hat{P}_\mathrm{Narain}$ of an asymmetric 
$\Z{2}\times\Z{2}$ orbifold. Note that the Narain point group is not isomorphic to 
$D_8\subset\mathrm{Sp}(4,\Z{})$ because in $\mathrm{Sp}(4,\Z{})$ we have 
$M_{(\mathrm{S}^2,\Id_2)} \neq M_{(\Id_2,\mathrm{S}^2)}$, while in 
$\mathrm{O}_{\hat{\eta}}(2,2+16,\Z{})$ the identity $(\hat{K}_\mathrm{S})^2=(\hat{C}_\mathrm{S})^2$ 
holds. In other words, the dictionary eq.~\eqref{eq:SP4ZBasisElements} from $\mathrm{Sp}(4,\Z{})$ 
to $\mathrm{O}_{\hat{\eta}}(2,2+16,\Z{})$ is two-to-one, such that 
$D_8/\Z{2} \cong \Z{2}\times\Z{2}$ with a $\Z{2}$ quotient that ensures 
$(\hat{K}_\mathrm{S})^2=(\hat{C}_\mathrm{S})^2$. 

In order to leave the generalized metric invariant eq.~\eqref{eq:GeneralizedMetricInvariant}, the 
string geometry is constrained as follows
\begin{equation}\label{eq:Z2xZ2WithoutWL}
G_{11} ~=~ \alpha' \quad,\quad G_{12} ~=~ B_{12}\quad\mathrm{and}\quad a_1 ~=~ a_2 ~=~ 0\;.
\end{equation}
Hence, the moduli $(T,U,Z)$ of this asymmetric $\Z{2}\times\Z{2}$ orbifold read
\begin{equation}
Z ~=~ 0 \quad,\quad T ~=~ U\;, \quad\mathrm{where}\quad U ~=~ \frac{1}{\alpha'}\,\left(G_{12} + \I\,\sqrt{\alpha'\,G_{22}-G_{12}^2}\right)\;,
\end{equation}
so that only $G_{12}$ and $G_{22}$ are free, in agreement with $\tau_1=\tau_2$ and $\tau_3=0$.

The normalizer $N(H)$ is generated by four elements given by
\begin{equation}
N(H) ~=~ \langle \;M_{(\mathrm{S},\mathrm{S})}\;,\;M_{(\mathrm{T},\mathrm{T})}\;,\;M_{(\mathrm{S}^2,\Id_2)}\;,\;M_{\times}\;\rangle ~\subset~ \mathrm{Sp}(4,\Z{})\;.
\end{equation}
Here, compared to ref.~\cite{Ding:2020zxw}, we use $M_{(\mathrm{S},\mathrm{S})}$ as generator of 
$N(H)$ instead of $M_{(\mathrm{S}^3,\mathrm{S}^3)}$. Translated to 
$\mathrm{O}_{\hat{\eta}}(2,2+16,\Z{})$ this yields the modular group  
\begin{equation}
\mathcal{G}_\mathrm{modular} ~=~ \langle\;\hat{K}_\mathrm{S}\hat{C}_\mathrm{S}\;,\;\hat{K}_\mathrm{T}\hat{C}_\mathrm{T}\;,\;\left(\hat{K}_\mathrm{S}\right)^2\;,\;\hat{M}\;\rangle ~\cong~ \mathrm{PSL}(2,\Z{}) \times \Z{2} \times \Z{2}^{\hat{M}}\;,
\end{equation}
of the asymmetric $\Z{2}\times\Z{2}$ orbifold. Note that we find $\mathrm{PSL}(2,\Z{})$ instead of 
$\mathrm{SL}(2,\Z{})$ because $\hat{K}_\mathrm{S}\hat{C}_\mathrm{S}$ is of order 2 (instead of 4). 
Then, the modulus $T$ transforms nontrivially under $\mathrm{PSL}(2,\Z{})$,
\begin{equation}
T ~\xmapsto{\hat{K}_\mathrm{S}\,\hat{C}_\mathrm{S}}~ -\frac{1}{T} \qquad\mathrm{and}\qquad
T ~\xmapsto{\hat{K}_\mathrm{T}\,\hat{C}_\mathrm{T}}~ T + 1\;,
\end{equation}
while it is invariant under $(\hat{K}_\mathrm{S})^2$ and $\hat{M}$.

According to table~1 of ref.~\cite{Ding:2021iqp} the fixed point $\tau_\mathrm{f}$ with 
$\tau_1=\tau_2$ and $\tau_3=0$ allows for a \CP transformation given by
\begin{equation}
\hat\CP ~=~ \hat{\Sigma}_*\qquad\mathrm{with}\qquad \hat\CP\,\hat\Theta_i\,\hat\CP^{-1} = \hat\Theta_i^{-1} \;\;\mathrm{for}\;\;i\in\{1,2\}\;.
\end{equation}
Hence, $\hat\CP$ is an outer automorphism of this asymmetric $\Z{2}\times\Z{2}$ Narain space group. 
Using eq.~\eqref{eq:TrafoOfModuli} for the \CP-like transformation $\hat\CP$, we find that $Z=0$ is 
invariant, while the modulus $T=U$ transforms as
\begin{equation}
T ~\xmapsto{\hat\CP}~ -\bar{T}\;.
\end{equation}

\subsubsection[Asymmetric Z2 x Z2 orbifold with discrete Wilson line]{\boldmath Asymmetric $\Z{2}\times\Z{2}$ orbifold with discrete Wilson line\unboldmath}
\label{sec:Z2xZ2AsymmetricWL}

Consider the fixed point
\begin{equation}
\tau_\mathrm{f} ~=~ \begin{pmatrix}\tau_1&\nicefrac{1}{2}\\\nicefrac{1}{2}&\tau_1\end{pmatrix} ~\in~ \mathcal{H}_2\;.
\end{equation}
In this case, the stabilizer is given by $\bar{H} \cong D_8/\{\pm\Id_4\}$, where $D_8$ is generated 
by two $\mathrm{Sp}(4,\Z{})$ elements,
\begin{equation}
h_1 := \begin{pmatrix}
 0& 1& 0& 0\\
 1& 0& 0& 0\\
 0& 0& 0& 1\\
 0& 0& 1& 0
\end{pmatrix} = M_{\times} \quad\mathrm{and}\quad 
h_2 := \begin{pmatrix}
 0& 1&-1& 0\\
-1& 0& 0& 1\\
 0& 0& 0& 1\\
 0& 0&-1& 0
\end{pmatrix} = \M{-1}{0}\,M_{\times}\, M_{(\Id_2,\mathrm{S}^2)}\;.
\end{equation}
Using the dictionary eq.~\eqref{eq:SP4ZBasisElements}, we choose two associated Narain twists
\begin{equation}
\hat\Theta_1 ~:=~ \hat{M} \quad\mathrm{and}\quad \hat\Theta_2 ~:=~ \W{-1}{0}\,\hat{M}\,\left(\hat{C}_\mathrm{S}\right)^2\;.
\end{equation}
In $\mathrm{O}_{\hat{\eta}}(2,2+16,\Z{})$, the two Narain twists $\hat\Theta_1$ and $\hat\Theta_2$ 
generate a $\Z{2}\times\Z{2}$ Narain point group of an asymmetric $\Z{2}\times\Z{2}$ orbifold. In 
order to leave the generalized metric invariant eq.~\eqref{eq:GeneralizedMetricInvariant}, we have 
to set
\begin{equation}
G_{11} ~=~ \alpha' \quad,\quad G_{12} ~=~ B_{12}\quad,\quad a_1 ~=~ 0 \quad\mathrm{and}\quad a_2 ~=~ \nicefrac{-1}{2}\;.
\end{equation}
This is similar to eq.~\eqref{eq:Z2xZ2WithoutWL} but with a non-trivial Wilson line $a_2 \neq 0$. 
Then, the moduli read
\begin{equation}
Z ~=~ \nicefrac{1}{2} \quad,\quad T ~=~ U\;, \quad\mathrm{where}\quad U ~=~ \frac{1}{\alpha'}\,\left(G_{12} + \I\,\sqrt{\alpha'\,G_{22}-G_{12}^2}\right)\;,
\end{equation}
so that only $G_{12}$ and $G_{22}$ are free. This confirms the expectation for 
$\tau_\mathrm{f}$ in this situation.

In order to identify the modular group $\mathcal{G}_\mathrm{modular}$ of this orbifold, we consider 
the normalizer
\begin{eqnarray}
N(H) &=& \langle \;\M{-1}{0}\,M_{(\mathrm{S}^2,\Id_2)}\;,\;M_{\times}\,\M{0}{-2}\,M_{\times}\,\M{0}{1}\,M_{(\mathrm{S}\mathrm{T}^{-1}\mathrm{S}^2,\mathrm{S}\mathrm{T}^2\mathrm{S}\mathrm{T})}\,\M{-1}{3} \;,\;\nonumber\\
    &\phantom{=}&\phantom{\langle}\;\M{-1}{0}\,M_{(\mathrm{S}^2\mathrm{T}^{-1},\mathrm{T}^{-1})}\;\rangle ~\subset~ \mathrm{Sp}(4,\Z{})\;,
\end{eqnarray}
see ref.~\cite{Ding:2020zxw}. We use the dictionary eq.~\eqref{eq:SP4ZBasisElements} to translate 
this into $\mathrm{O}_{\hat{\eta}}(2,2+16,\Z{})$. We find
\begin{equation}
\mathcal{G}_\mathrm{modular} ~=~ \langle\;\hat{M}_1\;,\;\hat{M}_2\;,\;\,\hat{M}_3\;\rangle\;,
\end{equation}
where we have defined
\begin{subequations}
\begin{eqnarray}
\hat{M}_1 & := & \W{-1}{0}\,\left(\hat{K}_\mathrm{S}\right)^2 ~=~ \hat\Theta_1\,\hat\Theta_2\;,\\
\hat{M}_2 & := & \hat{M}\,\W{0}{-2}\,\hat{M}\,\W{0}{1} \hat{K}_\mathrm{S}\,\left(\hat{K}_\mathrm{T}\right)^{-1}\,\left(\hat{K}_\mathrm{S}\right)^2\,\hat{C}_\mathrm{S}\,\left(\hat{C}_\mathrm{T}\right)^2\,\hat{C}_\mathrm{S}\,\hat{C}_\mathrm{T}\,\W{-1}{3}\;,\\
\hat{M}_3 & := & \W{-1}{0}\,\left(\hat{K}_\mathrm{S}\right)^2\,\left(\hat{K}_\mathrm{T}\right)^{-1}\,\left(\hat{C}_\mathrm{T}\right)^{-1}\;.
\end{eqnarray}
\end{subequations}
Since these transformations are outer automorphisms of the Narain space group, they give rise to 
the modular group of this asymmetric $\Z{2}\times\Z{2}$ orbifold with discrete Wilson line. Using 
the generalized metric eq.~\eqref{eq:TrafoOfModuli}, we identify their action on the modulus $T$ 
and obtain
\begin{equation}
T ~\xmapsto{~\hat{M}_2~}~ -\frac{2\,T+3}{4\,T+2} \quad\mathrm{and}\quad T ~\xmapsto{~\hat{M}_3~}~ T - 1\;,
\end{equation}
while $\hat{M}_1$ leaves $T$ invariant.

Finally, for the fixed point $\tau_\mathrm{f}$ with $\tau_1=\tau_2$ and $\tau_3=\nicefrac{1}{2}$ we 
identify \CP from table~1 of ref.~\cite{Ding:2021iqp} as
\begin{equation}
\hat\CP ~=~ \W{-1}{0}\,\hat{\Sigma}_*\qquad\mathrm{with}\qquad \hat\CP\,\hat\Theta_i\,\hat\CP^{-1} = \hat\Theta_i^{-1} \;\;\mathrm{for}\;\;i\in\{1,2\}\;.
\end{equation}
Hence, $\hat\CP$ is an outer automorphism of this asymmetric $\Z{2}\times\Z{2}$ Narain space group. 
Using eq.~\eqref{eq:TrafoOfModuli}, it leaves $Z=\nicefrac{1}{2}$ invariant, while the modulus 
$T=U$ transforms as
\begin{equation}
T ~\xmapsto{\hat\CP}~ -\bar{T}\;.
\end{equation}

\subsubsection[Asymmetric S3 orbifold]{\boldmath Asymmetric $S_3$ orbifold\unboldmath}
\label{sec:S3IAsymmetric}

The next fixed point in the list of ref.~\cite{Ding:2020zxw} reads
\begin{equation}
\tau_\mathrm{f} ~=~ \begin{pmatrix}\tau_1&\nicefrac{\tau_1}{2}\\\nicefrac{\tau_1}{2}&\tau_1\end{pmatrix} ~\in~ \mathcal{H}_2\;.
\end{equation}
This point is stabilized by $\bar{H} \cong S_3$, generated by two elements
\begin{subequations}
\begin{eqnarray}
h_1 & := & \begin{pmatrix}
-1& 1& 0& 0\\
 0& 1& 0& 0\\
 0& 0&-1& 0\\
 0& 0& 1& 1
\end{pmatrix} ~=~ M_{\times}\,\M{0}{1}\,M_{\times}\, M_{(\Id_2,\mathrm{S}^2)}\;,\\
h_2 & := & \begin{pmatrix}
 1& 0& 0& 0\\
 1&-1& 0& 0\\
 0& 0& 1& 1\\
 0& 0& 0&-1
\end{pmatrix} ~=~ \M{0}{1}\,M_{(\mathrm{S}^2,\Id_2)}\;.
\end{eqnarray}
\end{subequations}
As a consequence, we choose two corresponding $\mathrm{O}_{\hat{\eta}}(2,2+16,\Z{})$ Narain twists
\begin{equation}
\hat\Theta_1 ~:=~ \hat{M}\,\W{0}{1}\,\hat{M}\,\left(\hat{C}_\mathrm{S}\right)^2 \quad\mathrm{and}\quad
\hat\Theta_2 ~:=~ \W{0}{1}\,\left(\hat{K}_\mathrm{S}\right)^2\;.
\end{equation}
In $\mathrm{O}_{\hat{\eta}}(2,2+16,\Z{})$, the two Narain twists $\hat\Theta_1$ and $\hat\Theta_2$ 
satisfy the condition
\begin{equation}
\left(\hat\Theta_1\right)^2 ~=~ \left(\hat\Theta_2\right)^2 ~=~  \left(\hat\Theta_1\,\hat\Theta_2\right)^3 ~=~ \Id_{20}\;.
\end{equation}
Hence, they generate an $S_3$ Narain point group $\hat{P}_\mathrm{Narain}$ of an asymmetric $S_3$ 
orbifold. In order to leave the generalized metric invariant eq.~\eqref{eq:GeneralizedMetricInvariant}, 
we have to set
\begin{equation}\label{eq:GBAForS31}
G_{11} ~=~ \frac{3\,\alpha'}{4} \quad,\quad G_{12} ~=~ B_{12}\quad,\quad a_1 ~=~ \nicefrac{1}{2} \quad\mathrm{and}\quad a_2 ~=~ 0\;.
\end{equation}
In this case, the moduli read
\begin{equation}
Z ~=~ \nicefrac{T}{2} \quad,\quad T ~=~ U\;,
\end{equation}
as expected for the given values of $\tau_\mathrm{f}$.

According to ref.~\cite{Ding:2020zxw} and using the $\mathrm{Sp}(4,\Z{})$ generators defined in 
ref.~\cite{Baur:2020yjl}, the normalizer $N(H)$ can be generated by four elements 
\begin{eqnarray}
N(H) & = & \langle \;M_{(\mathrm{S},\mathrm{S}^3)}\,M_{\times}\;,\;M_{(\mathrm{S}^3,\mathrm{S}^3)}\,\M{-1}{0}\,M_{(\mathrm{T}^{-2},\mathrm{T}^{-2})}\,M_{(\mathrm{S},\mathrm{S})}\;,\;\nonumber\\
& & \phantom{\langle}\; M_{\times}\;,\;\M{0}{1}\,M_{(\mathrm{S}^2,\Id_2)}\;\rangle ~\subset~ \mathrm{Sp}(4,\Z{})\;.
\end{eqnarray}
Using the dictionary eq.~\eqref{eq:SP4ZBasisElements}, the corresponding group in 
$\mathrm{O}_{\hat{\eta}}(2,2+16,\Z{})$ reads
\begin{equation}
\mathcal{G}_\mathrm{modular} ~=~ \Big\langle\; \hat{M}_1\;,\;\hat{M}_2\;,\; \hat{M}\;,\;\W{0}{1}\left(\hat{K}_\mathrm{S}\right)^2\;\Big\rangle\;,
\end{equation}
where we defined
\begin{equation}
\hat{M}_1 ~:=~ \hat{K}_\mathrm{S}\left(\hat{C}_\mathrm{S}\right)^3\hat{M} \quad\mathrm{and}\quad \hat{M}_2 ~:=~ \left(\hat{K}_\mathrm{S}\right)^3\left(\hat{C}_\mathrm{S}\right)^3\W{-1}{0}\left(\hat{K}_\mathrm{T}\right)^{-2} \left(\hat{C}_\mathrm{T}\right)^{-2}\hat{K}_\mathrm{S}\hat{C}_\mathrm{S}\;.
\end{equation}
As the group $\mathcal{G}_\mathrm{modular}$ consists of outer automorphisms of the asymmetric $S_3$ 
Narain space group, this yields the modular group of this orbifold. Then, using 
eq.~\eqref{eq:TrafoOfModuli} the K\"ahler modulus $T$ transforms under the generators of 
$\mathcal{G}_\mathrm{modular}$ as
\begin{equation}
T ~\xmapsto{~\hat{M}_1~}~ -\frac{4}{3T} \quad\mathrm{and}\quad
T ~\xmapsto{~\hat{M}_2}~ \frac{2T}{3T+2}\;,
\end{equation}
while $T$ is invariant under $\hat{M}$ and $\W{0}{1}(\hat{K}_\mathrm{S})^2$.

Finally, for the fixed point $\tau_\mathrm{f}$ with $\tau_1=\tau_2$ and $\tau_3=\nicefrac{\tau_1}{2}$ 
we translate \CP from table~1 of ref.~\cite{Ding:2021iqp} into 
$\mathrm{O}_{\hat{\eta}}(2,2+16,\Z{})$, yielding
\begin{equation}
\hat\CP ~=~ \W{-1}{0}\,\hat{\Sigma}_*\qquad\mathrm{with}\qquad \hat\CP\,\hat\Theta_i\,\hat\CP^{-1} = \hat\Theta_i^{-1} \;\;\mathrm{for}\;\;i\in\{1,2\}\;.
\end{equation}
Consequently, $\hat\CP$ is an outer automorphism of this asymmetric $S_3$ Narain space group. We 
use eq.~\eqref{eq:TrafoOfModuli} to show that $\hat\CP$ acts as
\begin{equation}
T ~\xmapsto{\hat\CP}~ -\bar{T}\;.
\end{equation}

\subsection{Orbifolds with moduli spaces of dimension 0}

Finally, there are six inequivalent subspaces of complex dimension 0 that are left invariant by 
subgroups of $\mathrm{Sp}(4,\Z{})$~\cite{Ding:2020zxw}. As before, for each fixed point the 
corresponding orbifold is constructed by embedding the stabilizer $\bar{H}$ from $\mathrm{Sp}(4,\Z{})$ 
into $\mathrm{O}_{\hat{\eta}}(2,2+16,\Z{})$. This yields a Narain point group $\hat{P}_\mathrm{Narain}$ 
in $D=2$. In all cases discussed here, it turns out that $\hat{P}_\mathrm{Narain}$ gives rise to an 
asymmetric orbifold, whose moduli are stabilized geometrically, as expected. Note that the 
normalizers are given in terms of the stabilizers, $N(H)=H$, as there are no free moduli. Hence, 
embedding $N(H)$ into $\mathrm{O}_{\hat{\eta}}(2,2+16,\Z{})$ just gives 
$\mathcal{G}_\mathrm{modular}\cong\hat{P}_\mathrm{Narain}$. If the $D=2$ orbifold constructed here 
is a subsector of a full $D=6$ orbifold, $\mathcal{G}_\mathrm{modular}$ yields a traditional flavor 
symmetry, as a transformation is called modular only if some modulus transforms nontrivially. In 
addition, we consider \CP from ref.~\cite{Ding:2021iqp} and confirm that these transformations are 
also unbroken in the corresponding string constructions. As the string moduli are stabilized 
geometrically, one cannot move in moduli space away from the \CP-conserving point. Hence, \CP 
cannot be broken spontaneously by the moduli in these cases.

\subsubsection[Asymmetric Z5 orbifold]{\boldmath Asymmetric $\Z{5}$ orbifold\unboldmath}
\label{sec:Z5Asymmetric}

The next fixed point in the Siegel upper half-plane is given by
\begin{equation}
\tau_\mathrm{f} ~=~ \begin{pmatrix}\zeta&\zeta+\zeta^{-2}\\\zeta+\zeta^{-2}&-\zeta^{-1}\end{pmatrix} ~\in~ \mathcal{H}_2\;,
\end{equation}
where $\zeta:=\exp\left(\nicefrac{2\pi\I}{5}\right)$. In this case, the stabilizer 
$\bar{H} \cong \Z{5}$ is generated by
\begin{equation}
h ~:=~ \begin{pmatrix}
 0&-1&-1&-1\\
 0& 0&-1& 0\\
 0& 0& 0&-1\\
 1& 0& 0& 1
\end{pmatrix} ~=~ M_{(\Id_2,\mathrm{T})}\,M_{\times}\, \M{0}{1}\,M_{(\mathrm{S}^2,\mathrm{S}^3)}\;,
\end{equation}
using the $\mathrm{Sp}(4,\Z{})$ generators of ref.~\cite{Baur:2020yjl}. Then, we define the 
corresponding Narain twist $\hat\Theta\in\mathrm{O}_{\hat{\eta}}(2,2+16,\Z{})$
\begin{equation}
\hat\Theta ~:=~ \hat{C}_\mathrm{T}\,\hat{M}\,\W{0}{1}\,\left(\hat{K}_\mathrm{S}\right)^2\,\left(\hat{C}_\mathrm{S}\right)^3\;.
\end{equation}
Note that $h^5=-\Id_4$, which is identified with $+\Id_4$ in $\bar{H}$. Also the Narain twist 
$\hat\Theta$ is of order 5 such that $\hat\Theta$ defines a $\Z{5}$ Narain point group of an 
asymmetric $\Z{5}$ orbifold. This orbifold action is compatible with the generalized metric 
eq.~\eqref{eq:GeneralizedMetricInvariant} if
\begin{subequations}
\begin{align}
G_{11} &~=~ \frac{\alpha'}{2}\left(-5+3\,\sqrt{5}\right)\;, & G_{22} &~=~ G_{11}\;,                            & G_{12} &~=~ \frac{\alpha'}{2}\left(5-2\,\sqrt{5}\right)\;,\\
B_{12} &~=~ \frac{\alpha'}{2}\left(2-\sqrt{5}\right)\;,     & a_1    &~=~ \frac{1}{2}\left(3-\sqrt{5}\right)\;,& a_2    &~=~ \frac{1}{2}\left(-1+\sqrt{5}\right)\;.
\end{align}
\end{subequations}
Consequently, all moduli are fixed by the orbifold action to the values
\begin{equation}\label{eq:ModuliZ5}
T ~=~ -\zeta^{-1} \quad,\quad U ~=~ \zeta \quad\mathrm{and}\quad Z ~=~ \zeta + \zeta^{-2}\;,
\end{equation}
in agreement with the fixed point $\tau_\mathrm{f} \in \mathcal{H}_2$ of $\mathrm{Sp}(4,\Z{})$.

Finally, we translate the \CP generator of this case from table~1 of ref.~\cite{Ding:2021iqp} into 
$\mathrm{O}_{\hat{\eta}}(2,2+16,\Z{})$ and obtain
\begin{equation}
\hat\CP ~=~ \left(\left(\hat{K}_\mathrm{S}\,\hat{C}_\mathrm{S}\,\W{-1}{0}\right)^3\,\W{-1}{0}\right)^{-1}\,\hat{\Sigma}_*\qquad\mathrm{with}\qquad \hat\CP\,\hat\Theta\,\hat\CP^{-1} = \hat\Theta^{-1}\;.
\end{equation}
Furthermore, the geometrically stabilized moduli $(T,U,Z)$ are invariant under this \CP 
transformation. Its a trivial fact that \CP cannot be broken spontaneously by the moduli of this 
$\Z{5}$ orbifold sector since all moduli are completely fixed according to eq.~\eqref{eq:ModuliZ5}.

\subsubsection[Asymmetric S4 orbifold]{\boldmath Asymmetric $S_4$ orbifold\unboldmath}
\label{sec:S4Asymmetric}

Taking the fixed point
\begin{equation}
\tau_\mathrm{f} ~=~ \begin{pmatrix}\tilde\eta&\frac{1}{2}(\tilde\eta-1)\\\frac{1}{2}(\tilde\eta-1)&\tilde\eta\end{pmatrix} ~\in~ \mathcal{H}_2\;,
\end{equation}
where $\tilde\eta:=\frac{1}{3}(1+2\,\sqrt{2}\,\I)$, the stabilizer $\bar{H} \cong S_4$ is generated by 
two elements
\begin{equation}
h_1 ~:=~ \begin{pmatrix}
 0& 1& 0& 0\\
 1& 0& 0& 0\\
 0& 0& 0& 1\\
 0& 0& 1& 0
\end{pmatrix} \quad\mathrm{and}\quad
h_2 ~:=~ \begin{pmatrix}
-1& 1& 1& 0\\
 1& 0& 0& 1\\
-1& 0& 0& 0\\
 1&-1& 0& 1
\end{pmatrix}\;.
\end{equation}
We write these $\mathrm{Sp}(4,\Z{})$ elements in terms of the generators of ref.~\cite{Baur:2020yjl}
\begin{equation}
h_1 ~=~ M_{\times} \quad\mathrm{and}\quad h_2 ~=~ M_{(\Id_2,\mathrm{S}^3)}\,M_{\times}\, \M{0}{-1}\,M_{(\mathrm{S},\Id_2)}\M{-1}{-1}\;.
\end{equation}
Then, we define the corresponding Narain twists
\begin{equation}
\hat\Theta_1 ~:=~ \hat{M} \quad\mathrm{and}\quad \hat\Theta_2 ~:=~ \left(\hat{C}_\mathrm{S}\right)^3\,\hat{M}\,\W{0}{-1}\,\hat{K}_\mathrm{S}\,\W{-1}{-1}\;.
\end{equation}
Note that $(h_2)^4=-\Id_4$ is of order 8 in $H$ (and of order 4 in $\bar{H}$). The corresponding 
Narain twist $\hat\Theta_2$ is also of order 4. $\hat\Theta_1$ and $\hat\Theta_2$ obey the 
relations
\begin{equation}
\left(\hat\Theta_1\right)^2 ~=~ \left(\hat\Theta_2\right)^4 ~=~ \left(\hat\Theta_1\,\hat\Theta_2\right)^3 ~=~ \Id_{20}\;.
\end{equation}
Hence, $\hat\Theta_1$ and $\hat\Theta_2$ generate a $S_4$ Narain point group of an asymmetric 
$S_4$ orbifold. This orbifold action constrains the generalized metric 
eq.~\eqref{eq:GeneralizedMetricInvariant} such that
\begin{equation}
G_{11} ~=~ G_{22} ~=~ \frac{3\alpha'}{4}\;\;,\;\; G_{12} ~=~ \frac{\alpha'}{4}\;\;,\;\; B_{12} ~=~ \frac{\alpha'}{2}\;\;,\;\;a_1 ~=~ a_2 ~=~ \nicefrac{1}{2}\;.
\end{equation}
Consequently, all moduli of this orbifold are stabilized geometrically,
\begin{equation}\label{eq:ModuliS4}
T ~=~ U ~=~ \tilde\eta \quad\mathrm{and}\quad Z ~=~ \frac{1}{2}\left(\tilde\eta-1\right)\;.
\end{equation}
Hence, the subspace $\tau_1=\tau_2=\tilde\eta$ and $\tau_3=\frac{1}{2}(\tilde\eta-1)$ can be 
implemented in string theory using an asymmetric $S_4$ orbifold.

We obtain a \CP transformation for this asymmetric $S_4$ orbifold by translating the corresponding 
case from table~1 of ref.~\cite{Ding:2021iqp} into $\mathrm{O}_{\hat{\eta}}(2,2+16,\Z{})$. This 
yields 
\begin{equation}
\hat\CP ~=~ \hat{K}_\mathrm{S}\,\hat{C}_\mathrm{S}\,\W{-1}{0}\,\left(\hat{C}_\mathrm{T}\,\hat{K}_\mathrm{T}\right)^{-1}\hat{K}_\mathrm{S}\,\hat{C}_\mathrm{S}\,\hat{\Sigma}_*\qquad\mathrm{with}\qquad \hat\CP\,\hat\Theta_i\,\hat\CP^{-1} ~=~ \hat\Theta_i^{-1}\;,
\end{equation}
for $i\in\{1,2\}$. Then, we confirm that the geometrically stabilized string moduli $(T,U,Z)$ given 
in eq.~\eqref{eq:ModuliS4} are invariant under this \CP transformation.


\subsubsection[Asymmetric (Z4 x Z2) rtimes Z2 orbifold]{\boldmath Asymmetric $(\Z{4} \times\Z{2}) \rtimes \Z{2}$ orbifold\unboldmath}
\label{sec:Z4Z2Z2Asymmetric}

The fixed point
\begin{equation}
\tau_\mathrm{f} ~=~ \begin{pmatrix}\I&0\\0&\I\end{pmatrix} ~\in~ \mathcal{H}_2\;,
\end{equation}
is stabilized by $\bar{H} \cong (\Z{4} \times\Z{2}) \rtimes \Z{2}$, generated by three elements
\begin{equation}
h_1 := \begin{pmatrix}
 0& 0& 1& 0\\
 0&-1& 0& 0\\
-1& 0& 0& 0\\
 0& 0& 0&-1
\end{pmatrix} \;\mathrm{,}\;\;
h_2 := \begin{pmatrix}
 0& 0&-1& 0\\
 0& 0& 0& 1\\
 1& 0& 0& 0\\
 0&-1& 0& 0
\end{pmatrix} \;\;\mathrm{and}\;\;
h_3 := \begin{pmatrix}
 0& 1& 0& 0\\
 1& 0& 0& 0\\
 0& 0& 0& 1\\
 0& 0& 1& 0
\end{pmatrix}\;.
\end{equation}
We write these $\mathrm{Sp}(4,\Z{})$ elements in terms of the generators of ref.~\cite{Baur:2020yjl}
\begin{equation}
h_1 ~=~ M_{(\mathrm{S}^2,\mathrm{S})} \quad\mathrm{and}\quad h_2 ~=~ M_{(\mathrm{S},\mathrm{S}^3)} \quad\mathrm{and}\quad h_3 ~=~ M_{\times}\;.
\end{equation}
Then, we define the corresponding Narain twists
\begin{equation}
\hat\Theta_1 ~:=~ \left(\hat{K}_\mathrm{S}\right)^2\,\hat{C}_\mathrm{S} \quad\mathrm{and}\quad \hat\Theta_2 ~:=~ \hat{K}_\mathrm{S}\,\left(\hat{C}_\mathrm{S}\right)^3 \quad\mathrm{and}\quad \hat\Theta_3 ~:=~ \hat{M}\;.
\end{equation}
These Narain twists obey the relations
\begin{subequations}
\begin{align}
\left(\hat\Theta_1\right)^4 & ~=~ \left(\hat\Theta_2\right)^2 ~=~ \left(\hat\Theta_3\right)^2 ~=~ \Id_{20}\;,\\
\hat\Theta_1\,\hat\Theta_2 & ~=~ \hat\Theta_2\,\hat\Theta_1\quad,\quad\hat\Theta_2\,\hat\Theta_3 ~=~ \hat\Theta_3\,\hat\Theta_2\quad\mathrm{and}\quad\hat\Theta_3\,\hat\Theta_1\,\hat\Theta_3 ~=~ \hat\Theta_1\,\hat\Theta_2\;.
\end{align}
\end{subequations}
Hence, they define a $(\Z{4} \times\Z{2}) \rtimes \Z{2}$ Narain point group of an asymmetric 
$(\Z{4} \times\Z{2}) \rtimes \Z{2}$ orbifold. This orbifold action fixes all components of the 
generalized metric eq.~\eqref{eq:GeneralizedMetricInvariant} as follows
\begin{equation}
G_{11} ~=~ G_{22} ~=~ \alpha'\quad,\quad G_{12} ~=~ B_{12} ~=~ 0 \quad\mathrm{and}\quad a_1 ~=~ a_2 ~=~ 0\;.
\end{equation}
Consequently, all moduli are fixed
\begin{equation}\label{eq:ModuliZ4Z2Z2}
T ~=~ U ~=~ \I \quad\mathrm{and}\quad Z ~=~ 0\;,
\end{equation}
in agreement with $\tau_1=\tau_2=\I$ and $\tau_3=0$.

Finally, we consider the fixed point $\tau_\mathrm{f}$ with $\tau_1=\tau_2=\I$ and $\tau_3=0$ in 
table~1 of ref.~\cite{Ding:2021iqp} and translate the corresponding \CP transformation into 
$\mathrm{O}_{\hat{\eta}}(2,2+16,\Z{})$. Hence, we confirm explicitly that the outer automorphism of 
the Narain space group
\begin{equation}
\hat\CP ~=~ \hat{\Sigma}_*\qquad\mathrm{with}\qquad \hat\CP\,\hat\Theta_i\,\hat\CP^{-1} ~=~ \hat\Theta_i^{-1}\;,
\end{equation}
for $i\in\{1,2,3\}$, leaves the geometrically stabilized moduli $(T,U,Z)$ given in 
eq.~\eqref{eq:ModuliZ4Z2Z2} invariant. Hence, we have identified a \CP-like transformation of this 
orbifold theory.

\subsubsection[Asymmetric S3 x Z6 orbifold]{\boldmath Asymmetric $S_3\times\Z{6}$ orbifold\unboldmath}
\label{sec:S3xZ6Asymmetric}

Next, we consider the fixed point
\begin{equation}
\tau_\mathrm{f} ~=~ \begin{pmatrix}\omega&0\\0&\omega\end{pmatrix} ~\in~ \mathcal{H}_2\;.
\end{equation}
Its stabilizer $\bar{H} \cong S_3 \times \Z{6}$ is generated by three elements
\begin{equation}
h_1 := \begin{pmatrix}
 0& 0& 0&-1\\
 1& 0& 1& 0\\
 0& 1& 0& 1\\
-1& 0& 0& 0
\end{pmatrix} \;\mathrm{,}\;\;
h_2 := \begin{pmatrix}
 0& 1& 0& 0\\
 1& 0& 0& 0\\
 0& 0& 0& 1\\
 0& 0& 1& 0
\end{pmatrix}\;\;\mathrm{and}\;\;
h_3 := \begin{pmatrix}
 0& 0& 1& 0\\
 0& 0& 0&-1\\
-1& 0&-1& 0\\
 0& 1& 0& 1
\end{pmatrix}\;.
\end{equation}
$h_1$ and $h_2$ generate $S_3$, while $h_3$ is the generator of $\Z{6}$. These 
$\mathrm{Sp}(4,\Z{})$ elements can be written in terms of the generators of 
ref.~\cite{Baur:2020yjl} and we obtain
\begin{equation}
h_1 ~=~ M_{(\mathrm{T}^{-1}\mathrm{S},\mathrm{S}^3\mathrm{T})}\,M_{\times} \quad,\quad h_2 ~=~ M_{\times} \quad\mathrm{and}\quad h_3 ~=~ M_{(\mathrm{S}^3\mathrm{T},\mathrm{S}\mathrm{T})}\;.
\end{equation}
Using the dictionary eq.~\eqref{eq:SP4ZBasisElements}, we can define the corresponding Narain twists
\begin{equation}
\hat\Theta_1 ~:=~ \left(\hat{K}_\mathrm{T}\right)^{-1} \hat{K}_\mathrm{S}\,\left(\hat{C}_\mathrm{S}\right)^3 \hat{C}_\mathrm{T}\,\hat{M} \quad,\quad 
\hat\Theta_2 ~:=~ \hat{M} \quad\mathrm{and}\quad 
\hat\Theta_3 ~:=~ \left(\hat{K}_\mathrm{S}\right)^3 \hat{K}_\mathrm{T}\,\hat{C}_\mathrm{S}\,\hat{C}_\mathrm{T}\;.
\end{equation}
The Narain twists $\hat\Theta_1$ and $\hat\Theta_2$ satisfy the relations
\begin{equation}
\left(\hat\Theta_1\right)^2 ~=~ \left(\hat\Theta_2\right)^2 ~=~ \left(\hat\Theta_1\,\hat\Theta_2\right)^3 ~=~ \Id_{20}\;.
\end{equation}
Hence, they generate the permutation group $S_3$. Furthermore, the order 6 Narain twist 
$\hat\Theta_3$ commutes with both, $\hat\Theta_1$ and $\hat\Theta_2$. Consequently, the three 
Narain twists $\hat\Theta_1$, $\hat\Theta_2$ and $\hat\Theta_3$ generate an $S_3 \times \Z{6}$ 
Narain point group $\hat{P}_\mathrm{Narain}$, which we use to define an asymmetric $S_3\times\Z{6}$ 
orbifold. Next, we construct the Narain lattice of this orbifold by demanding invariance of the 
generalized metric eq.~\eqref{eq:GeneralizedMetricInvariant} under the three Narain twists. This 
yields
\begin{equation}
G_{11} ~=~ G_{22} ~=~ \alpha'\;\;,\;\; G_{12} ~=~ B_{12} ~=~ -\frac{\alpha'}{2}\;\;,\;\;a_1 ~=~ a_2 ~=~ 0\;.
\end{equation}
So, we see that all moduli $(T,U,Z)$ are stabilized and their values read
\begin{equation}\label{eq:ModuliForS3xZ6}
T ~=~ U ~=~ \omega \quad\mathrm{and}\quad Z~=~ 0\;.
\end{equation}
Thus, the invariant subspace $\tau_1=\tau_2=\omega$ and $\tau_3=0$ discussed in the bottom-up 
construction of ref.~\cite{Ding:2020zxw} can be constructed explicitly in string theory using an 
asymmetric orbifold compactification of $D=2$ dimensions with Narain point group 
$\hat{P}_\mathrm{Narain} \cong S_3 \times \Z{6}$.

Using table~1 of ref.~\cite{Ding:2021iqp} we identify \CP as
\begin{equation}
\hat\CP ~=~ \left(\hat{C}_\mathrm{T}\,\hat{K}_\mathrm{T}\right)^{-1}\,\hat{\Sigma}_* \qquad\mathrm{with}\qquad \hat\CP\,\hat\Theta_i\,\hat\CP^{-1} ~\in~ [\hat\Theta_i^{-1}]\;.
\end{equation}
Hence, $\hat\CP$ is a class-inverting outer automorphism~\cite{Chen:2014tpa} of 
$\hat{S}_\mathrm{Narain}$ that leaves the geometrically stabilized moduli 
eq.~\eqref{eq:ModuliForS3xZ6} invariant.

\subsubsection[Asymmetric S3 x Z2 orbifold]{\boldmath Asymmetric $S_3\times\Z{2}$ orbifold\unboldmath}
\label{sec:S3IIAsymmetric}

The next fixed point in the Siegel upper half-plane from the list of ref.~\cite{Ding:2020zxw} is 
given by
\begin{equation}
\tau_\mathrm{f} ~=~ \frac{\I}{\sqrt{3}}\begin{pmatrix}2&1\\1&2\end{pmatrix} ~\in~ \mathcal{H}_2\;.
\end{equation}
In this case, the stabilizer $\bar{H} \cong S_3 \times \Z{2}$ is generated by three elements
\begin{equation}
h_1 := \begin{pmatrix}
 0& 0& 0& 1\\
 0& 0& 1& 1\\
 1&-1& 0& 0\\
-1& 0& 0& 0
\end{pmatrix} \;\mathrm{,}\;\;
h_2 := \begin{pmatrix}
 0& 0& 1& 1\\
 0& 0& 1& 0\\
 0&-1& 0& 0\\
-1& 1& 0& 0
\end{pmatrix}\;\;\mathrm{and}\;\;
h_3 := \begin{pmatrix}
 0& 0& 0& 1\\
 0& 0&-1& 0\\
 0&-1& 0& 0\\
 1& 0& 0& 0
\end{pmatrix}\;.
\end{equation}
$h_1$ and $h_2$ generate $S_3$ (using that $(h_1)^2=(h_2)^2=-\Id_4$ is identified with $+\Id_4$ in 
$\bar{H}$), while the $\Z{2}$ factor is generated by $h_3$. These $\mathrm{Sp}(4,\Z{})$ elements 
can be decomposed in terms of the generators of ref.~\cite{Baur:2020yjl} and we find
\begin{equation}
h_1 ~=~ M_{\times}\,M_{(\mathrm{S},\mathrm{S})}\,\M{0}{-1} \quad,\quad 
h_2 ~=~ M_{(\mathrm{S},\mathrm{S})}\,\M{0}{-1}\,M_{\times} \quad\mathrm{and}\quad 
h_3 ~=~ M_{(\mathrm{S}^3,\mathrm{S})}\,M_{\times}\;.
\end{equation}
Next, we map these $\mathrm{Sp}(4,\Z{})$ elements into the Narain construction using 
ref.~\cite{Baur:2020yjl} and define the following Narain twists
\begin{equation}
\hat\Theta_1 ~:=~ \hat{M}\,\hat{K}_\mathrm{S}\,\hat{C}_\mathrm{S}\,\W{0}{-1}\quad,\quad 
\hat\Theta_2 ~:=~ \hat{K}_\mathrm{S}\,\hat{C}_\mathrm{S}\,\W{0}{-1}\,\hat{M} \quad\mathrm{and}\quad 
\hat\Theta_3 ~:=~ \left(\hat{K}_\mathrm{S}\right)^3 \hat{C}_\mathrm{S}\,\hat{M}\;.
\end{equation}
The Narain twists $\hat\Theta_1$ and $\hat\Theta_2$ satisfy the relations
\begin{equation}
\left(\hat\Theta_1\right)^2 ~=~ \left(\hat\Theta_2\right)^2 ~=~ \left(\hat\Theta_1\,\hat\Theta_2\right)^3 ~=~ \Id_{20}\;.
\end{equation}
Hence, they generate the permutation group $S_3$. Furthermore, the order 2 Narain twist 
$\hat\Theta_3$ commutes with both, $\hat\Theta_1$ and $\hat\Theta_2$. Thus, the Narain twists 
$\hat\Theta_1$, $\hat\Theta_2$ and $\hat\Theta_3$ generate an $S_3\times\Z{2}$ Narain point group. 
The resulting asymmetric $S_3\times\Z{2}$ orbifold restricts the generalized metric 
eq.~\eqref{eq:GeneralizedMetricInvariant} to a unique form, given by
\begin{equation}
G_{11} ~=~ \frac{3\,\alpha'}{4} \;\;,\;\; G_{22} ~=~ \alpha'\;\;,\;\; G_{12} ~=~ B_{12} ~=~ 0\;\;,\;\;a_1 ~=~ \nicefrac{1}{2} \;\;,\;\;a_2 ~=~ 0\;.
\end{equation}
This orbifold is similar to the $S_3$ orbifold with moduli given in eq.~\eqref{eq:GBAForS31}, but 
with the additional constraints $G_{22} = \alpha'$ and $B_{12} = 0$. As a consequence, the moduli 
$(T,U,Z)$ have to take the values
\begin{equation}
T ~=~ U ~=~ \frac{2\,\I}{\sqrt{3}} \quad\mathrm{and}\quad Z~=~ \frac{\I}{\sqrt{3}}\;,
\end{equation}
as expected from the bottom-up discussion in ref.~\cite{Ding:2020zxw}.

Also in this case, we can use table~1 of ref.~\cite{Ding:2021iqp} to identify a \CP-like 
transformation of the asymmetric $S_3\times\Z{2}$ orbifold,
\begin{equation}
\hat\CP ~=~ \hat{\Sigma}_* \qquad\mathrm{with}\qquad \hat\CP\,\hat\Theta_i\,\hat\CP^{-1} = \hat\Theta_i^{-1}\;,
\end{equation}
for all generators $i\in\{1,2,3\}$ of this $S_3\times\Z{2}$ Narain point group.

\subsubsection[Asymmetric Z12 orbifold]{\boldmath Asymmetric $\Z{12}$ orbifold\unboldmath}
\label{sec:Z12Asymmetric}

Finally, we consider the fixed point
\begin{equation}
\tau_\mathrm{f} ~=~ \begin{pmatrix}\omega&0\\0&\I\end{pmatrix} ~\in~ \mathcal{H}_2\;.
\end{equation}
This point is stabilized by $\bar{H}\cong\Z{12}$, which is generated by an element 
$h\in\mathrm{Sp}(4,\Z{})$ that we can write in terms of the generators defined in 
ref.~\cite{Baur:2020yjl} as follows
\begin{equation}
h ~:=~ \begin{pmatrix}
 0& 0& 1& 0\\
 0& 0& 0& 1\\
-1& 0&-1& 0\\
 0&-1& 0& 0
\end{pmatrix} ~=~ M_{(\mathrm{S},\mathrm{S}\mathrm{T})}\;.
\end{equation}
This $\mathrm{Sp}(4,\Z{})$ element can be mapped to a Narain twist 
$\hat\Theta\in\mathrm{O}_{\hat{\eta}}(2,2+16,\Z{})$ using the dictionary 
eq.~\eqref{eq:SP4ZBasisElements}, and we obtain
\begin{equation}
\hat\Theta ~:=~ \hat{K}_\mathrm{S}\,\hat{C}_\mathrm{S}\,\hat{C}_\mathrm{T}\;,
\end{equation}
which is of order 12. Hence, $\hat\Theta$ generates a $\Z{12}$ Narain point group and, 
consequently, an asymmetric $\Z{12}$ orbifold in $D=2$ dimensions, cf.\ ref.~\cite{Harvey:1987da} 
and section~8 of ref.~\cite{GrootNibbelink:2017usl}. The generalized metric $\mathcal{H}$ needs to 
be invariant, see eq.~\eqref{eq:GeneralizedMetricInvariant}, which fixes $\mathcal{H}$ to 
\begin{equation}
G_{11} ~=~ G_{22} ~=~ \frac{2\,\alpha'}{\sqrt{3}} \;\;,\;\; G_{12} ~=~ -\frac{\alpha'}{\sqrt{3}}\;\;,\;\; B_{12} ~=~ 0\;\;,\;\;a_1 ~=~ a_2 ~=~ 0\;.
\end{equation}
As a consequence, the moduli $(T,U,Z)$ have to take the values
\begin{equation}\label{eq:ModuliZ12}
T ~=~ \I \quad,\quad U ~=~ \omega \quad\mathrm{and}\quad Z ~=~ 0\;.
\end{equation}
Thus, we have found an explicit realization of the invariant subspace $\tau_1=U=\omega$, $\tau_2=T=\I$ 
and $\tau_3=Z=0$ in terms of an asymmetric $\Z{12}$ orbifold, where the moduli $(T,U,Z)=(\I,\omega,0)$ 
are frozen by the orbifold action.

Finally, we translate the \CP transformation from table~1 of ref.~\cite{Ding:2021iqp} for the case 
$\tau_1=U=\omega$, $\tau_2=T=\I$ and $\tau_3=Z=0$ into $\mathrm{O}_{\hat{\eta}}(2,2+16,\Z{})$. We 
find
\begin{equation}
\hat\CP ~=~ \left(\hat{C}_\mathrm{T}\right)^{-1}\,\hat{\Sigma}_* \qquad\mathrm{with}\qquad \hat\CP\,\hat\Theta\,\hat\CP^{-1} = \hat\Theta^{-1}\;.
\end{equation}
Thus, $\hat\CP$ is an outer automorphism of the Narain space group of this asymmetric $\Z{12}$ 
orbifold. As expected, it leaves the moduli eq.~\eqref{eq:ModuliZ12} invariant.

\section{Conclusion and Outlook}
\label{sec:Conclusions}

In the present work we have initiated the discussion of flavor symmetries of the Siegel modular 
group $\Sp{4}$ from a top-down perspective. In string theory, $\Sp{4}$ describes properties of the 
moduli $T$ and $U$ of a two-torus compactification as well as a Wilson line $Z$, as can be 
derived from the Narain lattice construction. This can be visualized through the moduli of a 
Riemann surface of genus $2$, which in the case of a vanishing Wilson line splits in two separate 
tori describing the $T$ and $U$ moduli independently.

The road to understand the relevance of $\Sp{4}$ for flavor symmetries of the Standard Model 
requires several steps. In a first step, we have to insist on the presence of chiral matter fields, 
which can be achieved by an orbifold twist. In the case of the previously discussed two-torus 
with vanishing Wilson lines, we had identified the possible orbifolds as those with twists $\Z{K}$ 
and $K=2,3,4,6$, with fixed points of the complex structure modulus $U$ at the boundaries of the 
fundamental domain of $\mathrm{SL}(2,\Z{})_U$. The generalization to $\Sp{4}$ then requires the 
classification of those orbifolds that lead to the fixed surfaces of $\Sp{4}$ in the Siegel upper 
half plane. These include two surfaces of complex dimension 2, five of complex dimension 1 and six 
of dimension 0. We have identified the 13 corresponding orbifolds explicitly and summarize our 
results in tables~\ref{tab:OrbifoldSummary} and~\ref{tab:OrbifoldSummaryLong}. In contrast to the 
previously discussed cases, we often find asymmetric orbifolds, which appear, for example, once we 
mod out the mirror symmetry (which interchanges $T$ and $U$). For each orbifold, we obtain the 
unbroken modular group $\mathcal{G}_\mathrm{modular}$ including \CP and the associated moduli 
transformations. With these results, we have completed the first step towards the understanding of 
the flavor structure of $\Sp{4}$.

In a second step, we would then have to analyze the properties of these orbifolds in detail. The 
symmetric orbifolds can be understood easily as they have a simple geometric interpretation. They 
extend the previously discussed cases $\Z{K}$ with $K=2,3,4,6$. The construction in 
section~\ref{sec:Z2Symmetric} corresponds to the $\Z{2}$ orbifold with complex moduli $T$ an $U$ 
and vanishing Wilson line. The cases with $K=3,4,6$ require a fixing of the complex structure 
modulus $U$, which is addressed in section~\ref{sec:Z4Symmetric} for $\Z{4}$ and 
section~\ref{sec:Z6Symmetric} for $\Z{3}$ and $\Z{6}$. The case discussed in 
section~\ref{sec:Z2Asymmetric} corresponds to an asymmetric orbifold with two complex moduli $T=U$ 
and Wilson line $Z$. As a direct geometric interpretation is lost in this case, an understanding of 
its properties requires further investigations. To regain the standard geometric picture it can be 
mapped to a symmetric $\Z{2}$ orbifold with moduli $T$ and $U$ and a quantized Wilson line 
$Z=\nicefrac{1}{2}$ as shown in section~\ref{sec:Z2SymmetricWL}. This reinterpretation of an 
asymmetric orbifold as a symmetric orbifold with specifically transformed moduli is not always 
possible. Some of the other asymmetric orbifolds of table~\ref{tab:OrbifoldSummary} cannot be 
mapped to symmetric orbifolds and need a more detailed analysis. Further studies should include the 
consideration of these asymmetric orbifolds towards the construction of models with the particle 
content of the Standard Model of particle physics.

\enlargethispage{\baselineskip}
In a third step, one would have to discuss for each orbifold the discrete flavor symmetries in the 
full eclectic picture of refs.~\cite{Nilles:2020nnc,Nilles:2020kgo}. This includes the traditional 
flavor group as well as the unbroken finite Siegel modular group that originates from the unbroken 
subgroup $\mathcal{G}_\mathrm{modular}$ of $\Sp{4}$ as given in section~\ref{sec:StabilizedModuliByOrb}. 
The finite Siegel modular groups are denoted by $\Gamma_{2,N}=\Sp{4}/\Gamma_2(N)$, where 
$\Gamma_2(N)$ is the principal congruence subgroup of $\Sp{4}$ with genus 2 and level $N$. This 
generalizes the homogeneous finite modular groups $\Gamma_{1,N}=\Gamma^\prime_N$ of 
$\mathrm{SL}(2,\Z{})$, discussed for example in ref.~\cite{Liu:2019khw}. For $N=2$ we have 
$\Gamma_{2,2}=\mathrm{Sp}(4,2)\cong S_6$ of order 720, while $\Gamma_{2,3}=\mathrm{Sp}(4,3)$
has already 51,840 elements~\cite{Conway:1985xxx}. These groups are huge, but they are usually 
not fully realized because of the orbifold twist that was introduced to get chiral matter from the 
torus. The task here would be to determine the unbroken finite Siegel modular groups for 
the 13 orbifold cases of table~\ref{tab:OrbifoldSummary} given the modular transformations 
determined in section~\ref{sec:StabilizedModuliByOrb}. This is beyond the scope of the present 
paper. Some clues can be found from the previously discussed cases with vanishing Wilson lines. The symmetric $\Z{2}$ 
orbifold from section~\ref{sec:Z2Symmetric} (with $Z=0$) is known to have a finite modular group 
$(S_3\times S_3)\rtimes \Z4$ (from $\Gamma^T_2\times \Gamma^U_2$ combined with mirror 
symmetry~\cite{Baur:2020jwc,Baur:2021mtl}) which nicely fits into $\Gamma_{2,2}=S_6$ for level 
$N=2$. It is not clear yet whether $\Gamma_{2,2}$ is also the relevant group for the asymmetric 
$\Z{2}$ orbifold discussed in section~\ref{sec:Z2Asymmetric}, although this seems plausible. 
On the other hand, in the case of the $\Z{3}$ orbifold with vanishing Wilson line, 
the finite modular group was found to be $T^\prime = \Gamma_3^\prime$ of level 
$N=3$~\cite{Lauer:1989ax,Lerche:1989cs,Lauer:1990tm,Baur:2019kwi,Baur:2019iai}. This leads to the 
conjecture that for the case described in section~\ref{sec:Z6Symmetric}, the finite modular group 
would descend from the finite Siegel modular group $\Gamma_{2,3}$, where we have confirmed that 
$\Gamma_{2,3}$ contains $T^\prime$. Thus, the result of this third step would be the 
determination of the finite Siegel modular flavor symmetry as well as the traditional flavor symmetry 
for each of the orbifolds given in table~\ref{tab:OrbifoldSummary}.

Once this has been achieved, the ultimate step to establish the full connection to bottom-up 
constructions is to determine the representations of the matter fields with respect to the full 
eclectic flavor group $\mathcal{G}_\mathrm{efg}$. Chiral fields tend to correspond to twisted 
fields located at the fixed points of the orbifold twist. In the $\Z{3}$ case, for example, we have 
three fixed points and matter fields transform as triplet representations of the traditional flavor 
symmetry $\Delta(54)\subset\mathcal{G}_\mathrm{efg}$ and as a $\rep{1} \oplus \rep{2}^\prime$ of 
the finite modular group $T^\prime\subset\mathcal{G}_\mathrm{efg}$. This shows, among others, that 
twisted fields need not correspond to irreducible representations of the finite modular group. So 
far, determining the representations of matter fields under the discrete flavor symmetries requires 
explicit computations of string vertex operators and their associated operator product 
expansions. This or the identification of a simpler method remains to be explored in detail for 
each of the orbifolds under consideration. This underlines that top-down model building with 
modular flavor symmetries has just begun to unfold its various possibilities. More work is 
needed in order to finally bridge the gap to bottom-up constructions.

The consideration of the finite Siegel modular flavor symmetry from a bottom-up perspective has 
been pioneered recently in ref.~\cite{Ding:2020zxw}. They considered the case with two 
unconstrained moduli: $T=U$ and a Wilson line $Z$. Superficially, this would look like 
case~\ref{sec:Z2Asymmetric} in our table~\ref{tab:OrbifoldSummary}, but this interpretation is not 
necessarily correct. By choosing the moduli at $T=U$ by hand, ref.~\cite{Ding:2020zxw} imposes 
$S_4\times\Z{2}$ as finite Siegel modular subgroup of $\Gamma_{2,2}\cong S_6$. In addition, some 
matter fields are postulated to build triplet representations of $S_4$. In our picture, the moduli 
can be stabilized at $T=U$ by an asymmetric $\Z2$ orbifold discussed in section~\ref{sec:Z2Asymmetric}. 
In principle, this orbifold also determines the unbroken finite Siegel modular group and the corresponding 
representations of matter fields. Unfortunately, their determination is technically involved and the 
results are currently not available. Hence, a detailed correspondence between ref.~\cite{Ding:2020zxw} 
and the top-down approach has yet to be clarified. To obtain a better geometric interpretation, one 
could reformulate this case as a symmetric orbifold with a quantized Wilson line as shown in 
section~\ref{sec:Z2SymmetricWL}. This might help to make contact to the discussion in the bottom-up 
approach of ref.~\cite{Ding:2020zxw}. We hope to report on the resolution of these questions in a 
future publication.

Finally, we stress that the results from our present endeavor may have interesting applications 
also in the study of other top-down scenarios. For example, in the context of magnetized toroidal 
compactifications~\cite{Kobayashi:2018bff,Ohki:2020bpo,Kikuchi:2020frp,Kikuchi:2020nxn,Hoshiya:2020hki,Kikuchi:2021ogn,Almumin:2021fbk} 
one typically derives the flavor properties of the models from the modular properties associated 
with the complex structure of a two-torus, disregarding the modular behavior of the K\"ahler 
and Wilson line moduli also present in the construction. It would be interesting to study how our 
considerations change the conclusions in these cases.

\section*{Acknowledgments}
P.V. is supported by the Deutsche Forschungsgemeinschaft (SFB1258).

\appendix

\section[Remark on mirror symmetry]{Remark on mirror symmetry} 
\label{app:mirror}

Note that in $\mathrm{Sp}(4,\Z{})$ the mirror transformation $M_{\times}$ can be expressed as
\begin{equation}\label{eq:SP4ZRelationForMirror}
M_{\times} ~=~ \M{0}{1}\,M_{(\mathrm{S},\Id_2)}\,\M{1}{0}\,M_{(\mathrm{S},\Id_2)}\,\M{0}{-1} ~\in~ \mathrm{Sp}(4,\Z{})\;.
\end{equation}
Let us denote the original definition from ref.~\cite{Baur:2020yjl} of mirror symmetry by $\hat{M}'$. 
Then, we use the dictionary eq.~\eqref{eq:SP4ZBasisElements} between $\mathrm{Sp}(4,\Z{})$ and 
$\mathrm{O}_{\hat{\eta}}(2,2+16,\Z{})$ and map the right-hand side of 
eq.~\eqref{eq:SP4ZRelationForMirror} into the modular group $\mathrm{O}_{\hat{\eta}}(2,2+16,\Z{})$ 
of the string setup and define 
\begin{equation}\label{eq:NewMirror}
\hat{M} ~:=~ \W{0}{1}\,\hat{K}_\mathrm{S}\,\W{1}{0}\,\hat{K}_\mathrm{S}\,\W{0}{-1} ~\in~ \mathrm{O}_{\hat{\eta}}(2,2+16,\Z{})\;.
\end{equation}
Crucially, the new mirror transformation $\hat{M}$ differs from the original definition of 
$\hat{M}'$,
\begin{equation}
\hat{M} ~\neq~ \hat{M}'\;.
\end{equation}
This is contrary to our expectation from eq.~\eqref{eq:SP4ZRelationForMirror}, as one would 
associate $M_{\times}\in\mathrm{Sp}(4,\Z{})$ with $\hat{M}'\in\mathrm{O}_{\hat{\eta}}(2,2+16,\Z{})$ 
using the dictionary eq.~\eqref{eq:SP4ZBasisElements} of ref.~\cite{Baur:2020yjl}. However, the 
generalized metric transforms identically under $\hat{M}$ and $\hat{M}'$, i.e.\ using e
q.~\eqref{eq:TrafoOfModuli} we find
\begin{subequations}
\begin{eqnarray}
\mathcal{H}(T,U,Z) & \xmapsto{\hat{M}\phantom{'}}  & \hat{M}^{-\mathrm{T}}\phantom{'}\mathcal{H}(T,U,Z)\, \hat{M}^{-1}\phantom{'} ~=~ \mathcal{H}(U,T,Z)\\
                   &                               & \quad\qquad\qquad\qquad\qquad\qquad\qquad\vequal\nonumber\\
\mathcal{H}(T,U,Z) & \xmapsto{\hat{M}'}            & \hat{M}'^{-\mathrm{T}}\mathcal{H}(T,U,Z)\, \hat{M}'^{-1} ~=~ \mathcal{H}(U,T,Z)\;.
\end{eqnarray}
\end{subequations}
Consequently, both transformations $\hat{M}$ and $\hat{M}'$ interchange $T$ and $U$ while leaving 
$Z$ invariant. Hence, on the level of the moduli, both transformations define mirror symmetry in 
the string construction. In the following, we use the new transformation
\begin{equation}\label{eq:NewMirrorMatrix}
\hat{M} ~=~ \begin{pmatrix}
 0& 0& 1& 0& 0& 0& 0\\
 0&-1& 0& 0& 0& 0& 0\\
 1& 0& 0& 0& 0& 0& 0\\
 0& 0& 0&-1& 0& 0& 0\\
 0& 0& 0& 0&-1& 1& 0\\
 0& 0& 0& 0& 0& 1& 0\\
 0& 0& 0& 0& 0& 0& \Id_{14}
\end{pmatrix}\;,
\end{equation}
defined in eq.~\eqref{eq:NewMirror} as the generator of mirror symmetry instead of $\hat{M}'$ that 
was defined in ref.~\cite{Baur:2020yjl}. Note that the new mirror transformation $\hat{M}$ also 
acts nontrivially on the 16 gauge degrees of freedom of the heterotic string as a $\Z{2}$ 
reflection.

\newpage
\section[Summary of results]{Summary of results} 
\label{app:results}

\bgroup
\setlength\LTleft{-1.5cm}
\setlength\LTright{-1.5cm}%
\begin{longtable}{ccl|ccll}
\caption{Stabilizing the moduli $(T,U,Z)$ at the fixed points of $\mathrm{Sp}(4,\Z{})$ by symmetric 
and asymmetric orbifold compactifications. We use the definitions 
$\omega :=\exp(\nicefrac{2\pi\I}{3})$, $\zeta:=\exp\left(\nicefrac{2\pi\I}{5}\right)$ and 
$\tilde\eta:=\frac{1}{3}(1+2\,\sqrt{2}\,\I)$.\label{tab:OrbifoldSummaryLong}}\\
\multicolumn{3}{c|}{\boldmath $\mathrm{Sp}(4,\Z{})$\unboldmath} & \multicolumn{4}{c}{\bf string theory}\\
\toprule
\endfirsthead
\caption{Stabilizing the moduli $(T,U,Z)$ at the fixed points of $\mathrm{Sp}(4,\Z{})$ by symmetric 
and asymmetric orbifold compactifications.}\\
\multicolumn{3}{c|}{\boldmath $\mathrm{Sp}(4,\Z{})$\unboldmath} & \multicolumn{4}{c}{\bf string theory}\\
\toprule
\endhead
\multicolumn{6}{r}{continued...}\\
\endfoot
\endlastfoot
                  &   &                                                                                                           & \multicolumn{4}{c}{\bf symmetric $\Z{2}$ orbifold}\\
\arrayrulecolor{lightgray}
\cmidrule{4-7}
\multirow{2}{*}{$\tau_\mathrm{f}$} & \multirow{2}{*}{=} & \multirow{2}{*}{$\begin{pmatrix}\tau_1&0\\0&\tau_2\end{pmatrix}$}       & $\mathcal{H}$             & : & \multicolumn{2}{l}{$a_1 = a_2 = 0$} \\
                  &   &                                                                                                           & moduli                    & : & \multicolumn{2}{l}{$(T,U,0)$} \\
\cmidrule{1-7} 
$\bar{H}$         & = & $\Big\langle\; M_{(\mathrm{S}^2,\Id_2)}\; \Big\rangle$                                                    & $\hat{P}_\mathrm{Narain}$ & = & \multicolumn{2}{l}{$\Big\langle\;\left(\hat{K}_\mathrm{S}\right)^2\;\Big\rangle ~\cong~ \Z{2}$}\\
\cmidrule{1-7} 
$N(H)$            & = & $\Big\langle \;M_{(\mathrm{S},\Id_2)}\;,\;M_{(\mathrm{T},\Id_2)}\;,\;$                                    &$\mathcal{G}_\mathrm{modular}$& = & \multicolumn{2}{l}{$\Big\langle\;\hat{K}_\mathrm{S}\;,\;\hat{K}_\mathrm{T}\;,\;\hat{C}_\mathrm{S}\;,\;\hat{C}_\mathrm{T}\;,\;\hat{M}\;\Big\rangle$}\\
                  &   & \;\;$M_{(\Id_2,\mathrm{S})}\;,\;M_{(\Id_2,\mathrm{T})}\;,\; M_{\times}\;\Big\rangle$                      &                           & $\cong$ & \multicolumn{2}{l}{$\left(\left(\SL{2,\Z{}}_T \times \SL{2,\Z{}}_U\right)/~\Z{2}\right) \rtimes \Z{2}^{\hat{M}}$}\\
\cmidrule{4-7} 
                  &   &                                                                                                           &                           &   & $T \xmapsto{\hat{K}_\mathrm{S}} -\frac{1}{T}$\;, & \!\!\!\!\!\!\!\!$U \xmapsto{\hat{K}_\mathrm{S}} U$\;,\\
                  &   &                                                                                                           &                           &   & $T \xmapsto{\hat{K}_\mathrm{T}} T+1$\;,          & \!\!\!\!\!\!\!\!$U \xmapsto{\hat{K}_\mathrm{T}} U$\;,\\
                  &   &                                                                                                           &                           &   & $T \xmapsto{\hat{C}_\mathrm{S}} T$\;,            & \!\!\!\!\!\!\!\!$U \xmapsto{\hat{C}_\mathrm{S}} -\frac{1}{U}$\;,\\
                  &   &                                                                                                           &                           &   & $T \xmapsto{\hat{C}_\mathrm{T}} T$\;,            & \!\!\!\!\!\!\!\!$U \xmapsto{\hat{C}_\mathrm{T}} U+1$\;,\\
                  &   &                                                                                                           &                           &   & $T \xmapsto{\hat{M}} U$\;,                       & \!\!\!\!\!\!\!\!$U \xmapsto{\hat{M}} T$\\
\cmidrule{1-7} 
$\CP_s$           & = & $\CP$                                                                                                     & $\hat\CP$                 & = & \multicolumn{2}{l}{$\hat{\Sigma}_*$}\\
\cmidrule{4-7} 
                  &   &                                                                                                           &                           &   & $T \xmapsto{\hat\CP} -\bar{T}$\;, & \!\!\!\!\!\!\!\!$U \xmapsto{\hat\CP} -\bar{U}$ \\
\arrayrulecolor{black}\midrule\arrayrulecolor{lightgray}
                  &   &                                                                                                           & \multicolumn{4}{c}{\bf asymmetric $\Z{2}$ orbifold}\\
\cmidrule{4-7}
\arrayrulecolor{lightgray}
\multirow{2}{*}{$\tau_\mathrm{f}$}&\multirow{2}{*}{=}&\multirow{2}{*}{$\begin{pmatrix}\tau_1&\tau_3\\\tau_3&\tau_1\end{pmatrix}$} & $\mathcal{H}$             & : & \multicolumn{2}{l}{$G_{11} = \alpha'(1-a_1^2)$\;,}\\
                  &   &                                                                                                           &                           &   & \multicolumn{2}{l}{$B_{12} = a_1\, a_2\, \alpha' + G_{12}$}\\
                  &   &                                                                                                           & moduli                    & : & \multicolumn{2}{l}{$(T,T,Z)$}\\
\cmidrule{1-7} 
$\bar{H}$         & = & $\Big\langle\; M_{\times}\; \Big\rangle$                                                                  & $\hat{P}_\mathrm{Narain}$ & = & \multicolumn{2}{l}{$\Big\langle\;\hat{M}\;\Big\rangle ~\cong~ \Z{2}$}\\
\cmidrule{1-7} 
$N(H)$            & = & $\Big\langle \;M_{(\mathrm{S},\mathrm{S})}\;,\;M_{(\mathrm{T},\mathrm{T})}\;,$                            &$\mathcal{G}_\mathrm{modular}$& = & \multicolumn{2}{l}{$\Big\langle\;\hat{K}_\mathrm{S}\,\hat{C}_\mathrm{S}\;,\;\hat{K}_\mathrm{T}\,\hat{C}_\mathrm{T}\;,\;\left(\hat{K}_\mathrm{S}\right)^2\;,\;\W{-1}{0}\;\Big\rangle$}\\
\cmidrule{4-7} 
                  &   & \;\;$M_{(\mathrm{S}^2,\Id_2)}\;,\;\M{-1}{0}\;\Big\rangle$                                                 &                           &   & $T \xmapsto{\hat{K}_\mathrm{S}\,\hat{C}_\mathrm{S}} -\frac{T}{T^2-Z^2}$\;, & \!\!\!\!\!\!\!\!$Z \xmapsto{\hat{K}_\mathrm{S}\,\hat{C}_\mathrm{S}} \frac{Z}{T^2-Z^2}$\;,\\
                  &   &                                                                                                           &                           &   & $T \xmapsto{\hat{K}_\mathrm{T}\,\hat{C}_\mathrm{T}} T + 1$\;,              & \!\!\!\!\!\!\!\!$Z \xmapsto{\hat{K}_\mathrm{T}\,\hat{C}_\mathrm{T}} Z$\;,\\
                  &   &                                                                                                           &                           &   & $T \xmapsto{\left(\hat{K}_\mathrm{S}\right)^2}      T$\;,                  & \!\!\!\!\!\!\!\!$Z \xmapsto{\left(\hat{K}_\mathrm{S}\right)^2}     -Z$\;,\\
                  &   &                                                                                                           &                           &   & $T \xmapsto{\W{-1}{0}}                              T$\;,                  & \!\!\!\!\!\!\!\!$Z \xmapsto{\W{-1}{0}}                              Z+1$\\
\cmidrule{1-7} 
$\CP_s$           & = & $\CP$                                                                                                     & $\hat\CP$                 & = & \multicolumn{2}{l}{$\hat{\Sigma}_*$}\\
\cmidrule{4-7} 
                  &   &                                                                                                           &                           &   & $T \xmapsto{\hat\CP} -\bar{T}$\;, & \!\!\!\!\!\!\!\!$Z \xmapsto{\hat\CP} -\bar{Z}$ \\
\arrayrulecolor{black}\midrule\arrayrulecolor{lightgray}
\pagebreak
                  &   &                                                                                                           & \multicolumn{4}{c}{\bf symmetric $\Z{2}$ orbifold with Wilson line}\\
\cmidrule{4-7}
\arrayrulecolor{lightgray}
\multirow{2}{*}{$\tau'_\mathrm{f}$}&\multirow{2}{*}{=}&\multirow{2}{*}{$\begin{pmatrix}\tau'_1&\nicefrac{1}{2}\\\nicefrac{1}{2}&\tau'_2\end{pmatrix}~=~b^{-1}\begin{pmatrix}\tau_1&\tau_3\\\tau_3&\tau_1\end{pmatrix}$}& $\mathcal{H}$             & : & \multicolumn{2}{l}{$a_1 = 0$\;,\; $a_2 = \nicefrac{-1}{2}$}\\
                  &   &                                                                                                           & moduli                    & : & \multicolumn{2}{l}{$(T,U,\nicefrac{1}{2})$}\\
\cmidrule{1-7} 
$\bar{H}$         & = & $\Big\langle\; M_{(\mathrm{S}^2,\Id_2)}\,\M{1}{0}\; \Big\rangle$                                          & $\hat{P}_\mathrm{Narain}$ & = & \multicolumn{2}{l}{$\Big\langle\;\left(\hat{C}_\mathrm{S}\right)^2\,\W{1}{0}\;\Big\rangle ~\cong~ \Z{2}$}\\
\cmidrule{1-7} 
$N(H)$            & = & $\Big\langle \;b^{-1}M_{(\mathrm{S},\mathrm{S})}\,b\;,\;b^{-1}M_{(\mathrm{T},\mathrm{T})}\,b\;,$          &$\mathcal{G}_\mathrm{modular}$& = & \multicolumn{2}{l}{$\Big\langle\;\hat{M}_1\;,\;\hat{M}_2\;,\;\hat{M}_3\;,\;\hat{M}_4\;\Big\rangle$}\\
                  &   & \;\;$b^{-1}M_{(\mathrm{S}^2,\Id_2)}\,b\;,\;b^{-1}\M{-1}{0}\,b\;\Big\rangle$                               &                           &   & \multicolumn{2}{l}{$\hat{M}_1 := \hat{B}^{-1}\,\hat{K}_\mathrm{S}\,\hat{C}_\mathrm{S}\,\hat{B}$\;,}\\ 
                  &   &                                                                                                           &                           &   & \multicolumn{2}{l}{$\hat{M}_2 := \hat{B}^{-1}\,\hat{K}_\mathrm{T}\,\hat{C}_\mathrm{T}\,\hat{B}$\;,}\\
                  &   &                                                                                                           &                           &   & \multicolumn{2}{l}{$\hat{M}_3 := \hat{B}^{-1}\,\left(\hat{K}_\mathrm{S}\right)^2\,\hat{B}$\;,}\\
                  &   &                                                                                                           &                           &   & \multicolumn{2}{l}{$\hat{M}_4 := \hat{B}^{-1}\,\W{-1}{0}\,\hat{B}$\;,}\\
                  &   & $b := M_{(\mathrm{T}^{-1},\Id_2)}\,\M{0}{1}\,M_{(\mathrm{T}\mathrm{S}^3,\Id_2)}$                          &                           &   & \multicolumn{2}{l}{$\hat{B} := \left(\hat{K}_\mathrm{T}\right)^{-1}\W{0}{1}\,\hat{K}_\mathrm{T}\,\left(\hat{K}_\mathrm{S}\right)^3$}\\
\cmidrule{4-7} 
                  &   &                                                                                                           &                           &   & $T \xmapsto{\hat{M}_1} -\frac{1}{4T}$\;,    & \!\!\!\!\!\!\!\!$U \xmapsto{\hat{M}_1} -\frac{1}{4U}$\;,\\
                  &   &                                                                                                           &                           &   & $T \xmapsto{\hat{M}_2} -\frac{T}{2\,T-1}$\;,& \!\!\!\!\!\!\!\!$U \xmapsto{\hat{M}_2} U + \frac{1}{2}$\;,\\
                  &   &                                                                                                           &                           &   & $T \xmapsto{\hat{M}_3} -\frac{1}{4U}$\;,    & \!\!\!\!\!\!\!\!$U \xmapsto{\hat{M}_3} -\frac{1}{4T}$\;,\\
                  &   &                                                                                                           &                           &   & $T \xmapsto{\hat{M}_4}  \frac{T}{2\,T+1}$\;,& \!\!\!\!\!\!\!\!$U \xmapsto{\hat{M}_4} U + \frac{1}{2}$\\
\cmidrule{1-7} 
$\CP_s$           & = & $M_{(\mathrm{S}^2,\Id_2)}\,\CP$                                                                           & $\hat\CP$                 & = & \multicolumn{2}{l}{$\left(\hat{K}_\mathrm{S}\right)^2\hat{\Sigma}_*$}\\
\cmidrule{4-7} 
                  &   &                                                                                                           &                           &   & $T \xmapsto{\hat\CP} -\bar{T}$\;, & \!\!\!\!\!\!\!\!$U \xmapsto{\hat\CP} -\bar{U}$ \\
\arrayrulecolor{black}\midrule\arrayrulecolor{lightgray}
                  &   &                                                                                                           & \multicolumn{4}{c}{\bf symmetric $\Z{4}$ orbifold}\\
\cmidrule{4-7}
\arrayrulecolor{lightgray}
\multirow{2}{*}{$\tau_\mathrm{f}$}&\multirow{2}{*}{=}&\multirow{2}{*}{$\begin{pmatrix}\I&0\\0&\tau_2\end{pmatrix}$}               & $\mathcal{H}$             & : & \multicolumn{2}{l}{$G_{11} = G_{22}$\;,\; $G_{12} = 0$\;,\; $a_1 = a_2 = 0$}\\
                  &   &                                                                                                           & moduli                    & : & \multicolumn{2}{l}{$(T,\I,0)$}\\
\cmidrule{1-7} 
$\bar{H}$         & = & $\Big\langle\; M_{(\Id_2,\mathrm{S})}\; \Big\rangle$                                                      & $\hat{P}_\mathrm{Narain}$ & = & \multicolumn{2}{l}{$\Big\langle\;\hat{C}_\mathrm{S}\;\Big\rangle ~\cong~ \Z{4}$}\\
\cmidrule{1-7} 
$N(H)$            & = & $\Big\langle \;M_{(\mathrm{S},\Id_2)}\;,\;M_{(\mathrm{T},\Id_2)}\;,\;M_{(\Id_2,\mathrm{S})}\;\Big\rangle$ &$\mathcal{G}_\mathrm{modular}$& = & \multicolumn{2}{l}{$\Big\langle\;\hat{K}_\mathrm{S}\;,\;\hat{K}_\mathrm{T}\;,\;\,\hat{C}_\mathrm{S}\;\Big\rangle$}\\
                  &   &                                                                                                           &                           & $\cong$ & \multicolumn{2}{l}{$\left(\SL{2,\Z{}}_T ~\times~ \Z{4}^R\right)/~\Z{2}$}\\
\cmidrule{4-7} 
                  &   &                                                                                                           &                           &   & $T \xmapsto{\hat{K}_\mathrm{S}} -\frac{1}{T}$\;, & \!\!\!\!\!\!\!\!$T \xmapsto{\hat{K}_\mathrm{T}} T+1$\;,\\
                  &   &                                                                                                           &                           &   & $T \xmapsto{\hat{C}_\mathrm{S}} T$ & \\
\cmidrule{1-7} 
$\CP_s$           & = & $\CP$                                                                                                     & $\hat\CP$                 & = & \multicolumn{2}{l}{$\hat{\Sigma}_*$}\\
\cmidrule{4-7} 
                  &   &                                                                                                           &                           &   & $T \xmapsto{\hat\CP} -\bar{T}$ & \\
\arrayrulecolor{black}\midrule\arrayrulecolor{lightgray}
\pagebreak
                  &   &                                                                                                           & \multicolumn{4}{c}{\bf symmetric $\Z{6}$ orbifold}\\
\cmidrule{4-7}
\multirow{2}{*}{$\tau_\mathrm{f}$}&\multirow{2}{*}{=}&\multirow{2}{*}{$\begin{pmatrix}\omega&0\\0&\tau_1\end{pmatrix}$}           & $\mathcal{H}$             & : & \multicolumn{2}{l}{$G_{11} = G_{22} = -2G_{12}$\;,\; $ a_1 = a_2 = 0$}\\
                  &   &                                                                                                           & moduli                    & : & \multicolumn{2}{l}{$(T,\omega,0)$}\\
\cmidrule{1-7} 
$\bar{H}$         & = & $\Big\langle\; M_{(\Id_2,\mathrm{S}^3\mathrm{T}\mathrm{S}\mathrm{T})}\;\Big\rangle$                       & $\hat{P}_\mathrm{Narain}$ & = & \multicolumn{2}{l}{$\Big\langle\;\left(\hat{C}_\mathrm{S}\right)^3\hat{C}_\mathrm{T}\,\hat{C}_\mathrm{S}\,\hat{C}_\mathrm{T}\;\Big\rangle ~\cong~ \Z{6}$}\\
\cmidrule{1-7} 
$N(H)$            & = &$\Big\langle\;M_{(\mathrm{S},\Id_2)}\;,\;M_{(\mathrm{T},\Id_2)}\;,\;M_{(\Id_2,\mathrm{S}^3\mathrm{T})}\;\Big\rangle$&$\mathcal{G}_\mathrm{modular}$& = & \multicolumn{2}{l}{$\Big\langle\;\hat{K}_\mathrm{S}\;,\;\hat{K}_\mathrm{T}\;,\;\left(\hat{C}_\mathrm{S}\right)^3\hat{C}_\mathrm{T}\;\Big\rangle$}\\
                  &   &                                                                                                           &                           & $\cong$ & \multicolumn{2}{l}{$\left(\SL{2,\Z{}}_T ~\times~ \Z{6}^R\right)/~\Z{2}$}\\
\cmidrule{4-7} 
                  &   &                                                                                                           &                           &   & \multicolumn{2}{l}{$T \xmapsto{\hat{K}_\mathrm{S}} -\frac{1}{T}$\;,\;$T \xmapsto{\hat{K}_\mathrm{T}} T+1$\;,\; $T \xmapsto{\left(\hat{C}_\mathrm{S}\right)^3\hat{C}_\mathrm{T}} T$}\\
\cmidrule{1-7} 
$\CP_s$           & = & $M_{(\Id_2,\mathrm{T}^{-1})}\,\CP$                                                                        & $\hat\CP$                 & = & \multicolumn{2}{l}{$\left(\hat{C}_\mathrm{T}\right)^{-1}\hat{\Sigma}_*$}\\
\cmidrule{4-7} 
                  &   &                                                                                                           &                           &   & $T \xmapsto{\hat\CP} -\bar{T}$ &\\
\arrayrulecolor{black}\midrule\arrayrulecolor{lightgray}
                  &   &                                                                                                           & \multicolumn{4}{c}{\bf asymmetric $\Z{2}\times\Z{2}$ orbifold}\\
\cmidrule{4-7}

\multirow{2}{*}{$\tau_\mathrm{f}$}&\multirow{2}{*}{=}&\multirow{2}{*}{$\begin{pmatrix}\tau_1&0\\0&\tau_1\end{pmatrix}$}           & $\mathcal{H}$             & : & \multicolumn{2}{l}{$G_{11} = \alpha'$\;,\; $G_{12} = B_{12}$\;,\; $a_1 = a_2 = 0$}\\
                  &   &                                                                                                           & moduli                    & : & \multicolumn{2}{l}{$(T,T,0)$}\\
\cmidrule{1-7} 
$\bar{H}$         & = & $\Big\langle\; M_{\times}\;,\;M_{\times}\,M_{(\mathrm{S}^2,\Id_2)}\;\Big\rangle$                          & $\hat{P}_\mathrm{Narain}$ & = & \multicolumn{2}{l}{$\Big\langle\;\hat{M}\;,\;\hat{M}\left(\hat{K}_\mathrm{S}\right)^2\;\Big\rangle ~\cong~ \Z{2}\times\Z{2}$}\\

\cmidrule{1-7} 
$N(H)$            & = &$\Big\langle\;M_{(\mathrm{S},\mathrm{S})}\;,\;M_{(\mathrm{T},\mathrm{T})}\;,\;$                            &$\mathcal{G}_\mathrm{modular}$& = & \multicolumn{2}{l}{$\Big\langle\;\hat{K}_\mathrm{S}\hat{C}_\mathrm{S}\;,\;\hat{K}_\mathrm{T}\hat{C}_\mathrm{T}\;,\;\left(\hat{K}_\mathrm{S}\right)^2\;,\;\hat{M}\;\Big\rangle$}\\
                  &   &\;\;$M_{(\mathrm{S}^2,\Id_2)}\;,\;M_{\times}\;\Big\rangle$                                                 &                           & $\cong$ & \multicolumn{2}{l}{$\mathrm{PSL}(2,\Z{}) \times \Z{2} \times \Z{2}^{\hat{M}}$}\\
\cmidrule{4-7} 
                  &   &                                                                                                           &                           &   & $T \xmapsto{\hat{K}_\mathrm{S}\,\hat{C}_\mathrm{S}} -\frac{1}{T}$\;, & \!\!\!\!\!\!\!\!$T \xmapsto{\hat{K}_\mathrm{T}\,\hat{C}_\mathrm{T}} T + 1$\;,\\
                  &   &                                                                                                           &                           &   & $T \xmapsto{\left(\hat{K}_\mathrm{S}\right)^2} T$\;,                 & \!\!\!\!\!\!\!\!$T \xmapsto{\hat{M}} T$\\
\cmidrule{1-7} 
$\CP_s$           & = & $\CP$                                                                                                     & $\hat\CP$                 & = & \multicolumn{2}{l}{$\hat{\Sigma}_*$}\\
\cmidrule{4-7} 
                  &   &                                                                                                           &                           &   & $T \xmapsto{\hat\CP} -\bar{T}$ & \\
\arrayrulecolor{black}\midrule\arrayrulecolor{lightgray}
                  &   &                                                                                                           & \multicolumn{4}{c}{\bf asymmetric $\Z{2}\times\Z{2}$ orbifold with Wilson line}\\
\cmidrule{4-7}
\multirow{2}{*}{$\tau_\mathrm{f}$}&\multirow{2}{*}{=}&\multirow{2}{*}{$\begin{pmatrix}\tau_1&\nicefrac{1}{2}\\\nicefrac{1}{2}&\tau_1\end{pmatrix}$}&$\mathcal{H}$&:& \multicolumn{2}{l}{$G_{11} = \alpha'$\;,\; $G_{12} = B_{12}$\;,\; $a_1 = 0$\;,\; $a_2 = \nicefrac{-1}{2}$}\\
                  &   &                                                                                                           & moduli                    & : & \multicolumn{2}{l}{$(T,T,\nicefrac{1}{2})$}\\
\cmidrule{1-7} 
$\bar{H}$         & = & $\Big\langle\; M_{\times}\;,\; \M{-1}{0}\,M_{\times}\, M_{(\Id_2,\mathrm{S}^2)}\;\Big\rangle$             & $\hat{P}_\mathrm{Narain}$ & = & \multicolumn{2}{l}{$\Big\langle\; \hat{M}\;,\;\W{-1}{0}\,\hat{M}\,\left(\hat{C}_\mathrm{S}\right)^2\;\Big\rangle ~\cong~ \Z{2}\times\Z{2}$}\\
\cmidrule{1-7} 
$N(H)$            & = & $\Big\langle \;\M{-1}{0}\,M_{(\mathrm{S}^2,\Id_2)}\;,\;$                                                  &$\mathcal{G}_\mathrm{modular}$& = & \multicolumn{2}{l}{$\Big\langle\;\hat{M}_1\;,\;\hat{M}_2\;,\;\,\hat{M}_3\;\Big\rangle$}\\
                  &   & \;$M_{\times}\,\M{0}{-2}\,M_{\times}\,\M{0}{1}\,\times$                                                   &                           &   & \multicolumn{2}{l}{$\hat{M}_1 :=  \W{-1}{0}\,\left(\hat{K}_\mathrm{S}\right)^2 = \hat\Theta_1\,\hat\Theta_2\;,$}\\
                  &   & \;$M_{(\mathrm{S}\mathrm{T}^{-1}\mathrm{S}^2,\mathrm{S}\mathrm{T}^2\mathrm{S}\mathrm{T})}\,\M{-1}{3}\;,\;\ldots$&                     &   & \multicolumn{2}{l}{$\hat{M}_2 := \hat{M}\,\W{0}{-2}\hat{M}\,\W{0}{1}\hat{K}_\mathrm{S}\,\left(\hat{K}_\mathrm{T}\right)^{-1}\times\ldots$}\\
\cmidrule{1-7} 
                  &   & \;$\ldots\M{-1}{0}\,M_{(\mathrm{S}^2\mathrm{T}^{-1},\mathrm{T}^{-1})}\;\Big\rangle$                       &                           &   & \multicolumn{2}{l}{$\phantom{\hat{M}_2 := }\ldots\left(\hat{K}_\mathrm{S}\right)^2\,\hat{C}_\mathrm{S}\,\left(\hat{C}_\mathrm{T}\right)^2\,\hat{C}_\mathrm{S}\,\hat{C}_\mathrm{T}\,\W{-1}{3}\;,$}\\
                  &   &                                                                                                           &                           &   & \multicolumn{2}{l}{$\hat{M}_3 := \W{-1}{0}\,\left(\hat{K}_\mathrm{S}\right)^2\,\left(\hat{K}_\mathrm{T}\right)^{-1}\,\left(\hat{C}_\mathrm{T}\right)^{-1}$}\\
\cmidrule{4-7} 
                  &   &                                                                                                           &                           &   & $T \xmapsto{\hat{M}_2} -\frac{2\,T+3}{4\,T+2}$\;, & \!\!\!\!\!\!\!\!$T \xmapsto{\hat{M}_3} T - 1$\;,\\
                  &   &                                                                                                           &                           &   & $T \xmapsto{\hat{M}_1} T$                         & \\
\cmidrule{1-7} 
$\CP_s$           & = & $\M{-1}{0}\,\CP$                                                                                          & $\hat\CP$                 & = & \multicolumn{2}{l}{$\W{-1}{0}\,\hat{\Sigma}_*$}\\
\cmidrule{4-7} 
                  &   &                                                                                                           &                           &   & $T \xmapsto{\hat\CP} -\bar{T}$ & \\
\arrayrulecolor{black}\midrule\arrayrulecolor{lightgray}
                  &   &                                                                                                           & \multicolumn{4}{c}{\bf asymmetric $S_3$ orbifold}\\
\cmidrule{4-7}
\multirow{2}{*}{$\tau_\mathrm{f}$}&\multirow{2}{*}{=}&\multirow{2}{*}{$\begin{pmatrix}\tau_1&\nicefrac{\tau_1}{2}\\\nicefrac{\tau_1}{2}&\tau_1\end{pmatrix}$}& $\mathcal{H}$ & : & \multicolumn{2}{l}{$G_{11} = \frac{3\alpha'}{4}$\;,\; $G_{12} = B_{12}$\;,\; $a_1 = \nicefrac{1}{2}$\;,\; $a_2 = 0$}\\
                  &   &                                                                                                           & moduli                    & : & \multicolumn{2}{l}{$(T,T,\nicefrac{T}{2})$}\\
\cmidrule{1-7} 
$\bar{H}$         & = & $\Big\langle\;M_{\times}\,\M{0}{1}\,M_{\times}\, M_{(\Id_2,\mathrm{S}^2)}\;,\;$                           & $\hat{P}_\mathrm{Narain}$ & = & \multicolumn{2}{l}{$\Big\langle\; \hat{M}\,\W{0}{1}\,\hat{M}\,\left(\hat{C}_\mathrm{S}\right)^2\;,\;\W{0}{1}\,\left(\hat{K}_\mathrm{S}\right)^2\;\Big\rangle$}\\
                  &   & \;\;$\M{0}{1}\,M_{(\mathrm{S}^2,\Id_2)}\;\Big\rangle$                                                     &                           &$\cong$& \multicolumn{2}{l}{$S_3$}\\
\cmidrule{1-7} 

$N(H)$            & = & $\Big\langle \;M_{(\mathrm{S},\mathrm{S}^3)}\,M_{\times}\;,\;$                                            &$\mathcal{G}_\mathrm{modular}$& = & \multicolumn{2}{l}{$\Big\langle\; \hat{M}_1\;,\;\hat{M}_2\;,\; \hat{M}\;,\;\W{0}{1}\left(\hat{K}_\mathrm{S}\right)^2\;\Big\rangle$}\\
                  &   & \;\;$M_{(\mathrm{S}^3,\mathrm{S}^3)} \M{-1}{0} M_{(\mathrm{T}^{-2},\mathrm{T}^{-2})} M_{(\mathrm{S},\mathrm{S})},$&                   &   & \multicolumn{2}{l}{$\hat{M}_1 := \hat{K}_\mathrm{S}\left(\hat{C}_\mathrm{S}\right)^3\hat{M}$}\\
                  &   & \;\;$M_{\times}\;,\;\M{0}{1}\,M_{(\mathrm{S}^2,\Id_2)}\;\Big\rangle$                                      &                           &   & \multicolumn{2}{l}{$\hat{M}_2 := \left(\hat{K}_\mathrm{S}\right)^3\left(\hat{C}_\mathrm{S}\right)^3\W{-1}{0}\left(\hat{K}_\mathrm{T}\right)^{-2}\times$}\\
                  &   &                                                                                                           &                           &   & \multicolumn{2}{l}{$\phantom{\hat{M}_2 := }\left(\hat{C}_\mathrm{T}\right)^{-2}\hat{K}_\mathrm{S}\hat{C}_\mathrm{S}$}\\
\cmidrule{4-7} 
                  &   &                                                                                                           &                           &   & $T \xmapsto{\hat{M}_1} -\frac{4}{3T}$\;, & \!\!\!\!\!\!\!\!$T \xmapsto{\hat{M}_2} \frac{2T}{3T+2}$\;,\\
                  &   &                                                                                                           &                           &   & $T \xmapsto{\hat{M}} T$\;,               & \!\!\!\!\!\!\!\!$T \xmapsto{\W{0}{1}\,\left(\hat{K}_\mathrm{S}\right)^2} T$\\
\cmidrule{1-7} 
$\CP_s$           & = & $\M{-1}{0}\,\CP$                                                                                          & $\hat\CP$                 & = & \multicolumn{2}{l}{$\W{-1}{0}\,\hat{\Sigma}_*$}\\
\cmidrule{4-7} 
                  &   &                                                                                                           &                           &   & $T \xmapsto{\hat\CP} -\bar{T}$ & \\
\arrayrulecolor{black}\midrule\arrayrulecolor{lightgray}
                  &   &                                                                                                           & \multicolumn{4}{c}{\bf asymmetric $\Z{5}$ orbifold}\\
\cmidrule{4-7}
\multirow{2}{*}{$\tau_\mathrm{f}$}&\multirow{2}{*}{=}&\multirow{2}{*}{$\begin{pmatrix}\zeta&\zeta+\zeta^{-2}\\\zeta+\zeta^{-2}&-\zeta^{-1}\end{pmatrix}$}& $\mathcal{H}$& : & $G_{11} = \frac{\alpha'}{2}\left(-5+3\,\sqrt{5}\right)$, & \!\!\!\!\!\!$B_{12} = \frac{\alpha'}{2}\left(2-\sqrt{5}\right)$,\\
                  &   &                                                                                                           &                           &   & $G_{12} = \frac{\alpha'}{2}\left(5-2\,\sqrt{5}\right)$\;,          & \!\!\!\!\!\!$a_1    = \frac{1}{2}\left(3-\sqrt{5}\right)$\;,\\
                  &   &                                                                                                           &                           &   & $G_{22} = G_{11}$\;,                                               & \!\!\!\!\!\!$a_2    = \frac{1}{2}\left(-1+\sqrt{5}\right)$\\
                  &   &                                                                                                           & moduli                    & : & \multicolumn{2}{l}{$(-\zeta^{-1}, \zeta, \zeta + \zeta^{-2})$}\\
\cmidrule{1-7} 
$\bar{H}$         & = & $\Big\langle\;M_{(\Id_2,\mathrm{T})}\,M_{\times}\, \M{0}{1}\,M_{(\mathrm{S}^2,\mathrm{S}^3)}\;\Big\rangle$& $\hat{P}_\mathrm{Narain}$ & = & \multicolumn{2}{l}{$\Big\langle\; \hat{C}_\mathrm{T}\,\hat{M}\,\W{0}{1}\,\left(\hat{K}_\mathrm{S}\right)^2\,\left(\hat{C}_\mathrm{S}\right)^3\;\Big\rangle~\cong~\Z{5}$}\\
\cmidrule{1-7} 
$\CP_s$           & = & $\left(\left(M_{(\mathrm{S},\mathrm{S})}\,\M{-1}{0}\right)^3\M{-1}{0}\right)^{-1}\CP$                     & $\hat\CP$                 & = & \multicolumn{2}{l}{$\left(\left(\hat{K}_\mathrm{S}\,\hat{C}_\mathrm{S}\,\W{-1}{0}\right)^3\,\W{-1}{0}\right)^{-1}\,\hat{\Sigma}_*$}\\
\arrayrulecolor{black}\midrule\arrayrulecolor{lightgray}
\pagebreak
                  &   &                                                                                                           & \multicolumn{4}{c}{\bf asymmetric $S_4$ orbifold}\\
\cmidrule{4-7}
\multirow{2}{*}{$\tau_\mathrm{f}$}&\multirow{2}{*}{=}&\multirow{2}{*}{$\begin{pmatrix}\tilde\eta&\frac{1}{2}(\tilde\eta-1)\\\frac{1}{2}(\tilde\eta-1)&\tilde\eta\end{pmatrix}$}& $\mathcal{H}$& : & \multicolumn{2}{l}{$G_{11} = G_{22} = \frac{3\alpha'}{4}$\;,\; $G_{12} = \frac{\alpha'}{4}$\;,}\\
                  &   &                                                                                                           &                           &   & \multicolumn{2}{l}{$B_{12} = \frac{\alpha'}{2}$\;,\; $a_1 = a_2 = \nicefrac{1}{2}$}\\
                  &   &                                                                                                           & moduli                    & : & \multicolumn{2}{l}{$(\tilde\eta, \tilde\eta, \frac{1}{2}\left(\tilde\eta-1\right))$}\\
\cmidrule{1-7} 
$\bar{H}$         & = & $\Big\langle\;M_{\times} \;,\; M_{(\Id_2,\mathrm{S}^3)}\,M_{\times}\, \M{0}{-1}\,\times$                  & $\hat{P}_\mathrm{Narain}$ & = & \multicolumn{2}{l}{$\Big\langle\; \hat{M} \;,\; \left(\hat{C}_\mathrm{S}\right)^3\,\hat{M}\,\W{0}{-1}\,\hat{K}_\mathrm{S}\,\W{-1}{-1}\;\Big\rangle$}\\
                  &   & \;\;$M_{(\mathrm{S},\Id_2)}\M{-1}{-1}\;\Big\rangle$                                                       &                           &$\cong$& \multicolumn{2}{l}{$S_4$}\\
\cmidrule{1-7} 
$\CP_s$           & = & $M_{(\mathrm{S},\mathrm{S})} \M{-1}{0} M_{(\mathrm{T}^{-1},\mathrm{T}^{-1})} M_{(\mathrm{S},\mathrm{S})} \CP$ & $\hat\CP$             & = & \multicolumn{2}{l}{$\hat{K}_\mathrm{S}\,\hat{C}_\mathrm{S}\,\W{-1}{0}\,\left(\hat{C}_\mathrm{T}\,\hat{K}_\mathrm{T}\right)^{-1}\hat{K}_\mathrm{S}\,\hat{C}_\mathrm{S}\,\hat{\Sigma}_*$}\\
\arrayrulecolor{black}\midrule\arrayrulecolor{lightgray}
                  &   &                                                                                                           & \multicolumn{4}{c}{\bf asymmetric $(\Z{4} \times\Z{2}) \rtimes \Z{2}$ orbifold}\\
\cmidrule{4-7}
\multirow{2}{*}{$\tau_\mathrm{f}$}&\multirow{2}{*}{=}&\multirow{2}{*}{$\begin{pmatrix}\I&0\\0&\I\end{pmatrix}$}                   & $\mathcal{H}$             & : & \multicolumn{2}{l}{$G_{11} = G_{22} = \alpha'$\;,\; $G_{12} = B_{12} = 0$\;,}\\
                  &   &                                                                                                           &                           &   & \multicolumn{2}{l}{$a_1 = a_2 = 0$}\\
                  &   &                                                                                                           & moduli                    & : & \multicolumn{2}{l}{$(\I, \I, 0)$}\\
\cmidrule{1-7} 
$\bar{H}$         & = & $\Big\langle\;M_{(\mathrm{S}^2,\mathrm{S})}\;,\;M_{(\mathrm{S},\mathrm{S}^3)}\;,\;M_{\times}\;\Big\rangle$& $\hat{P}_\mathrm{Narain}$ & = & \multicolumn{2}{l}{$\Big\langle\; \left(\hat{K}_\mathrm{S}\right)^2\,\hat{C}_\mathrm{S}\;,\;\hat{K}_\mathrm{S}\,\left(\hat{C}_\mathrm{S}\right)^3\;,\;\hat{M} \;\Big\rangle$}\\
                  &   &                                                                                                           &                           & $\cong$ & \multicolumn{2}{l}{$(\Z{4} \times\Z{2}) \rtimes \Z{2}$}\\
\cmidrule{1-7} 
$\CP_s$           & = & $\CP$                                                                                                     & $\hat\CP$                 & = & \multicolumn{2}{l}{$\hat{\Sigma}_*$}\\
\arrayrulecolor{black}\midrule\arrayrulecolor{lightgray}
                  &   &                                                                                                           & \multicolumn{4}{c}{\bf asymmetric $S_3\times\Z{6}$ orbifold}\\
\cmidrule{4-7}
\multirow{2}{*}{$\tau_\mathrm{f}$}&\multirow{2}{*}{=}&\multirow{2}{*}{$\begin{pmatrix}\omega&0\\0&\omega\end{pmatrix}$}           & $\mathcal{H}$             & : & \multicolumn{2}{l}{$G_{11} = G_{22} = \alpha'$\;,\; $G_{12} = B_{12} = -\frac{\alpha'}{2}$\;,}\\
                  &   &                                                                                                           &                           &   & \multicolumn{2}{l}{$a_1 = a_2 = 0$}\\
                  &   &                                                                                                           & moduli                    & : & \multicolumn{2}{l}{$(\omega, \omega, 0)$}\\
\cmidrule{1-7} 
$\bar{H}$         & = & $\Big\langle\;M_{(\mathrm{T}^{-1}\mathrm{S},\mathrm{S}^3\mathrm{T})}\,M_{\times}\;,\;M_{\times}\;,\;$     & $\hat{P}_\mathrm{Narain}$ & = & \multicolumn{2}{l}{$\Big\langle\; \left(\hat{K}_\mathrm{T}\right)^{-1} \hat{K}_\mathrm{S}\,\left(\hat{C}_\mathrm{S}\right)^3 \hat{C}_\mathrm{T}\,\hat{M}\;,\;\hat{M}\;,\;$}\\
                  &   & \;\;$M_{(\mathrm{S}^3\mathrm{T},\mathrm{S}\mathrm{T})}\;\Big\rangle$                                      &                           &   & \multicolumn{2}{l}{\;\;$\left(\hat{K}_\mathrm{S}\right)^3 \hat{K}_\mathrm{T}\,\hat{C}_\mathrm{S}\,\hat{C}_\mathrm{T} \;\Big\rangle ~\cong~ S_3 \times\Z{6}$}\\
\cmidrule{1-7} 
$\CP_s$           & = & $M_{(\mathrm{T}^{-1},\mathrm{T}^{-1})}\,\CP$                                                              & $\hat\CP$                 & = & \multicolumn{2}{l}{$\left(\hat{C}_\mathrm{T}\,\hat{K}_\mathrm{T}\right)^{-1}\,\hat{\Sigma}_*$}\\
\arrayrulecolor{black}\midrule\arrayrulecolor{lightgray}
                  &   &                                                                                                           & \multicolumn{4}{c}{\bf asymmetric $S_3\times\Z{2}$ orbifold}\\
\cmidrule{4-7}
\multirow{2}{*}{$\tau_\mathrm{f}$}&\multirow{2}{*}{=}&\multirow{2}{*}{$\frac{\I}{\sqrt{3}}\begin{pmatrix}2&1\\1&2\end{pmatrix}$}  & $\mathcal{H}$             & : & \multicolumn{2}{l}{$G_{11} = \frac{3\alpha'}{4}$\;,\; $G_{22} = \alpha'$\;,\; $G_{12} = B_{12} = 0$\;,}\\
                  &   &                                                                                                           &                           &   & \multicolumn{2}{l}{$a_1 = \nicefrac{1}{2}$\;,\; $a_2 = 0$}\\
                  &   &                                                                                                           & moduli                    & : & \multicolumn{2}{l}{$(\frac{2\I}{\sqrt{3}}, \frac{2\I}{\sqrt{3}}, \frac{\I}{\sqrt{3}})$}\\
\cmidrule{1-7} 
$\bar{H}$         & = & $\Big\langle\;M_{\times}\,M_{(\mathrm{S},\mathrm{S})}\,\M{0}{-1}\;,$                                      & $\hat{P}_\mathrm{Narain}$ & = & \multicolumn{2}{l}{$\Big\langle\;  \hat{M}\,\hat{K}_\mathrm{S}\,\hat{C}_\mathrm{S}\,\W{0}{-1}\;,\;\hat{K}_\mathrm{S}\,\hat{C}_\mathrm{S}\,\W{0}{-1}\,\hat{M}\;,\;$}\\
                  &   & $M_{(\mathrm{S},\mathrm{S})}\,\M{0}{-1}M_{\times}\;,\;M_{(\mathrm{S}^3,\mathrm{S})}\,M_{\times}\;\Big\rangle$&                        &   & \multicolumn{2}{l}{\;\;$\left(\hat{K}_\mathrm{S}\right)^3 \hat{C}_\mathrm{S}\,\hat{M} \;\Big\rangle ~\cong~ S_3 \times\Z{2}$}\\
\cmidrule{1-7} 
$\CP_s$           & = & $\CP$                                                                                                     & $\hat\CP$                 & = & \multicolumn{2}{l}{$\hat{\Sigma}_*$}\\
\arrayrulecolor{black}\midrule\arrayrulecolor{lightgray}
\pagebreak
                  &   &                                                                                                           & \multicolumn{4}{c}{\bf asymmetric $\Z{12}$ orbifold}\\
\cmidrule{4-7}
\multirow{2}{*}{$\tau_\mathrm{f}$}&\multirow{2}{*}{=}&\multirow{2}{*}{$\begin{pmatrix}\omega&0\\0&\I\end{pmatrix}$}               & $\mathcal{H}$             & : & \multicolumn{2}{l}{$G_{11} = G_{22} = \frac{2\alpha'}{\sqrt{3}}$\;,\; $G_{12} = -\frac{\alpha'}{\sqrt{3}}$\;,}\\
                  &   &                                                                                                           &                           &   & \multicolumn{2}{l}{$B_{12} = 0$\;,\; $a_1 = a_2 = 0$}\\
                  &   &                                                                                                           & moduli                    & : & \multicolumn{2}{l}{$(\I, \omega, 0)$}\\
\cmidrule{1-7} 
$\bar{H}$         & = & $\Big\langle\; M_{(\mathrm{S},\mathrm{S}\mathrm{T})} \;\Big\rangle$                                       & $\hat{P}_\mathrm{Narain}$ & = & \multicolumn{2}{l}{$\Big\langle\; \hat{K}_\mathrm{S}\,\hat{C}_\mathrm{S}\,\hat{C}_\mathrm{T}\;\Big\rangle ~\cong~ \Z{12}$}\\
\cmidrule{1-7} 
$\CP_s$           & = & $M_{(\Id_2,\mathrm{T}^{-1})}\,\CP$                                                                        & $\hat\CP$                 & = & \multicolumn{2}{l}{$\left(\hat{C}_\mathrm{T}\right)^{-1}\,\hat{\Sigma}_*$}\\

\arrayrulecolor{black}
\bottomrule
\end{longtable}
\egroup

\providecommand{\bysame}{\leavevmode\hbox to3em{\hrulefill}\thinspace}

\end{document}